\documentclass[twocolumn]{aastex61}
\usepackage{amsmath}
\usepackage{color}

\newcommand{\rom}[1]{\textup{\uppercase\expandafter{\romannumeral#1}}}

\newcommand{\nd}{\multicolumn{1}{c}{$\dots$}}
\newcommand{\mcr}[1]{\multicolumn{1}{r}{#1}}

\shorttitle{Leavitt laws for type \rom{2} Cepheid variables}
\shortauthors{Bhardwaj et al.}

\begin{document}
\title{Large Magellanic Cloud Near-Infrared Synoptic Survey. \rom{4}.\\Leavitt Laws for Type \rom{2} Cepheid Variables}

\correspondingauthor{Anupam Bhardwaj}
\email{anupam.bhardwajj@gmail.com, abhardwaj@eso.org}

\vspace{7pt}
\author{Anupam Bhardwaj}
\affiliation{Department of Physics \& Astrophysics, University of Delhi, Delhi 110007, India}
\affiliation{European Southern Observatory, Karl-Schwarzschild-Stra\ss e 2, 85748, Garching, Germany}
\vspace{7pt}
\author{~Lucas M.~Macri}
\affiliation{Mitchell Institute for Fundamental Physics \& Astronomy, Department of Physics \& Astronomy, Texas A\&M University, College Station, TX 77843, USA}
\vspace{5pt}
\author{~Marina Rejkuba}
\affiliation{European Southern Observatory, Karl-Schwarzschild-Stra\ss e 2, 85748, Garching, Germany}
\vspace{7pt}
\author{~Shashi M.~Kanbur}
\affiliation{State University of New York, Oswego, NY 13126, USA}
\vspace{7pt}
\author{~Chow-Choong Ngeow}
\affiliation{Graduate Institute of Astronomy, National Central University, Jhongli 32001, Taiwan}
\vspace{7pt}
\author{~Harinder P.~Singh}
\affiliation{Department of Physics \& Astrophysics, University of Delhi, Delhi 110007, India}

\received{}
\revised{}
\accepted{}
\vspace{20pt}

\begin{abstract}
We present time-series observations of Population \rom{2} Cepheids in the Large Magellanic Cloud at near-infrared ($JHK_s$) wavelengths. Our sample consists of 81 variables with accurate periods and optical ($VI$) magnitudes from the OGLE survey, covering various subtypes of pulsators (BL Herculis, W Virginis and RV Tauri). We generate light curve templates using high-quality $I$-band data in the LMC from OGLE and $K_s$-band data in the Galactic Bulge from VVV and use them to obtain robust mean magnitudes. We derive Period-Luminosity (P-L) relations in the near-infrared and Period-Wesenheit (P-W) relations by combining optical and near-infrared data. Our P-L and P-W relations are consistent with published work when excluding long-period RV Tauris. We find that Pop II Cepheids and RR Lyraes follow the same P-L relations in the LMC. Therefore, we use trigonometric parallax from the {\it Gaia DR1} for VY~Pyx and the {\it Hubble Space Telescope} parallaxes for $k$~Pav and 5 RR Lyrae variables to obtain an absolute calibration of the Galactic $K_s$-band P-L relation, resulting in a distance modulus to the LMC of $\mu_{\rm LMC} = 18.54\pm0.08$~mag. We update the mean magnitudes of Pop~II Cepheids in Galactic globular clusters using our light curve templates and obtain distance estimates to those systems, anchored to a precise late-type eclipsing binary distance to the LMC. We find the distances to these globular clusters based on Pop~II Cepheids are consistent (within $2\sigma$) with estimates based on the $M_V-[\rm{Fe}/\rm{H}]$ relation for horizontal branch stars.
\end{abstract}

\section{Introduction}
\label{sec:intro}

Classical Cepheid variables are Population~I stars used as standard candles for the extragalactic distance scale, thanks to their high luminosities and a well-defined Period-Luminosity relation (PLR) \citep[the ``Leavitt Law'',][]{leavitt}. They are the primary distance indicator used in the most accurate and precise determination of the Hubble constant to date \citep{riess16}. Type~II Cepheids (hereafter, T2Cs) are low-mass, Pop~II stars which can be found in globular clusters, disk, bulge and halo environments \citep{wallerstein2002, sandage2006}. Classical and T2Cs follow different PLRs, with the latter more than a magnitude fainter than the former at similar periods. T2Cs are further classified based on their periods as BL Herculis (BLH, $1\lesssim\!P\!\lesssim4$~d), W Virginis (WVI\footnote{We exclusively use ``WVI'' to refer to this subtype and not to the Wesenheit relation in the $VI$ bands.}, $4\lesssim\!P\!\lesssim20$~d) and RV Tauri (RVT, $P\!\gtrsim 20$~d). The classification of RVT is often ambiguous because they show irregular light curves, but they are considered a subtype of T2Cs \citep{sandage2006, feast2008}. \citet{soszynski2008a} suggested another subtype, peculiar W Virginis (PWV, $4\lesssim\!P\!\lesssim10$~d), with distinct light curves that are mostly brighter and bluer than WVI.

The PLRs of T2Cs at optical bands have been studied extensively \citep[][and references therein]{nemec1994, alcock1998, kubiak2003, majaess2009, schmidt2009}. These relations exhibit possible non-linearities, which coupled with fainter absolute magnitudes (relative to Classical Cepheids) limits their use as potential distance indicators. The third phase of the Optical Gravitational Lensing Experiment (OGLE-\rom{3}) presented optical light curves and PLRs for T2Cs in the Galactic Bulge and the Magellanic Clouds (MCs) in \citet{soszynski2008a, soszynski2010, soszynski2011}. They found that Bulge T2Cs are dominated by short-period BLH stars which are more luminous than their counterparts in the Clouds, and that LMC RVT stars lie above the PLR followed by the shorter-period classes. Theoretical studies of T2Cs based on pulsating models and evolutionary calculations by \citet{bono1997} found that their masses decrease with increasing period and they follow a Period-Luminosity-Amplitude relation in the $B$ band. 

Over the past decade, the increased availability of large-format and higher-quality near-infrared (hereafter, NIR) detectors has made it possible to study increasingly larger samples of Cepheids at longer wavelengths where PLRs are less sensitive to extinction and metallicity \citep{madore1991}. \citet{matsunaga2006, matsunaga2009, matsunaga2011} derived NIR PLRs for T2Cs in Galactic globular clusters (GGCs) and the MCs. The authors found evidence for different slopes in the PLRs of each system as well as a varying frequency of each subtype. \citet{feast2008, gmat2008, ciech2010} discussed the application of NIR PLRs of T2Cs and \citet{gmat2008} estimated a distance to the Galactic Center of $R_0=7.94\pm0.37$~kpc using these variables. Recently, \citet{ripepi2015} presented $JK_s$ observations of T2Cs in the MCs from the VMC survey \citep{cioni2011} and derived a variety of P-L, Period-Luminosity-Color (PLC) and Period-Wesenheit relations (PWR).

This paper is the fourth in a series of articles based on observations obtained by the Large Magellanic Cloud Near-infrared Synoptic Survey \citep[LMCNISS,][hereafter, Paper \rom{1}]{macri2015}. Paper I presented survey details and the absolute calibration of NIR PLRs for Classical Cepheids. \citet[][hereafter, Papers \rom{2}~and~\rom{3}]{bhardwaj2016a, bhardwaj2016b} derived PWRs for Classical Cepheids, studied possible non-linearities in the Leavitt Laws and estimated Cepheid-based distances to Local Group galaxies. In this paper we focus on the NIR observations of T2Cs and their corresponding relations.

The rest of the paper is structured as follows: \S\ref{sec:data} describes the observations, data reduction and photometry of T2Cs; \S\ref{sec:fou} discusses the variation of light-curve parameters as a function of period and wavelength and the construction of templates; \S4 contains the derivation of NIR PLRs and PWRs, a comparison to published work, and an estimate of the distance to the LMC; \S\ref{sec:ggcdist} presents template fits to observations of T2Cs in Galactic globular clusters and the resulting distance estimates; \S\ref{sec:discuss} summarizes our results.

\section{The Data}
\label{sec:data}

\citet[][Paper I]{macri2015} carried out a time-series survey of 18~sq.~deg. in the central region of the LMC at $JHK_s$ wavelengths using the 1.5-m telescope at the Cerro Tololo Interamerican Observatory and the CPAPIR camera. Observations were carried out in queue mode by the SMARTS consortium during 32 nights from Nov 2006 to Nov 2007. The survey products include measurements for more than $3.5\times10^6$ sources, including $\sim1500$ Classical Cepheids. Interested readers should refer to Paper I for details of the data reduction, time-series photometry, magnitude calibration and artificial star simulations used to derive crowding corrections. Given the fainter nature of T2Cs, their crowding corrections were more significant than those of Classical Cepheids and ranged from 0.001 mag to 0.10, 0.08 and 0.27 mag in $JHK_s$, respectively.

We cross-matched the LMCNISS catalog against OGLE-\rom{3} \citep{soszynski2008a} and identified 81 T2Cs with periods ranging from 1 to 68~d; 70 of these have $JHK_s$ measurements while the remaining 11 only have data in $J$ and/or $H$-band. The sample consists of 16 BLH, 31 WVI, 12 PWV and 22 RVT stars.  The NIR times-series photometry for these objects are presented in Table~\ref{table:phot}. We adopt the period ($P$), time of maximum brightness in $I$-band ($T_{I,\rm max}$) and optical ($VI$) mean magnitudes from OGLE-\rom{3}.

\begin{deluxetable}{lcrlll}
\tablewidth{0pt}
\tablecaption{ Time-series photometry for Type \rom{2} Cepheids. \label{table:phot}}
\tablehead{\colhead{ID} & \colhead{Band} & \colhead{MJD} & \colhead{Phase} & \colhead{mag} & \colhead{$\sigma$}}
\startdata
025&  $J$  &   42.620& 0.879& 13.633& 0.021\\
025&  $J$  &   42.753& 0.881& 13.609& 0.021\\
\nd&  \nd  &\nd	  &	\nd & \nd   & \nd  \\
025&  $H$  &   42.623& 0.879& 13.513& 0.018\\
025&  $H$  &   42.756& 0.881& 13.451& 0.040\\
\nd&  \nd  &\nd	  &	\nd & \nd   & \nd  \\
025&  $K_s$&   42.625& 0.879& 13.348& 0.015\\
025&  $K_s$&   42.758& 0.881& 13.370& 0.032\\
\nd&\nd    &\nd	     &	\nd & \nd   & \nd \\ 
\enddata
\tablecomments{ ID: OGLE-LMC-T2CEP-NNN, from OGLE-\rom{3} catalog of Type \rom{2} Cepheids \citep{soszynski2008a}; ${\rm MJD}={\rm JD}-2450000$; phase is determined using the period and time of maximum brightness in $I$-band from OGLE-\rom{3}. Fifth column represents magnitude in a near-infrared band and sixth column lists its associated uncertainty. The entire table is available online as supplemental material; sample time-series in $JHK_s$ for a T2Cs is shown here for guidance regarding its content.}
\end{deluxetable}

\begin{figure}[t]
\includegraphics[width=0.5\textwidth]{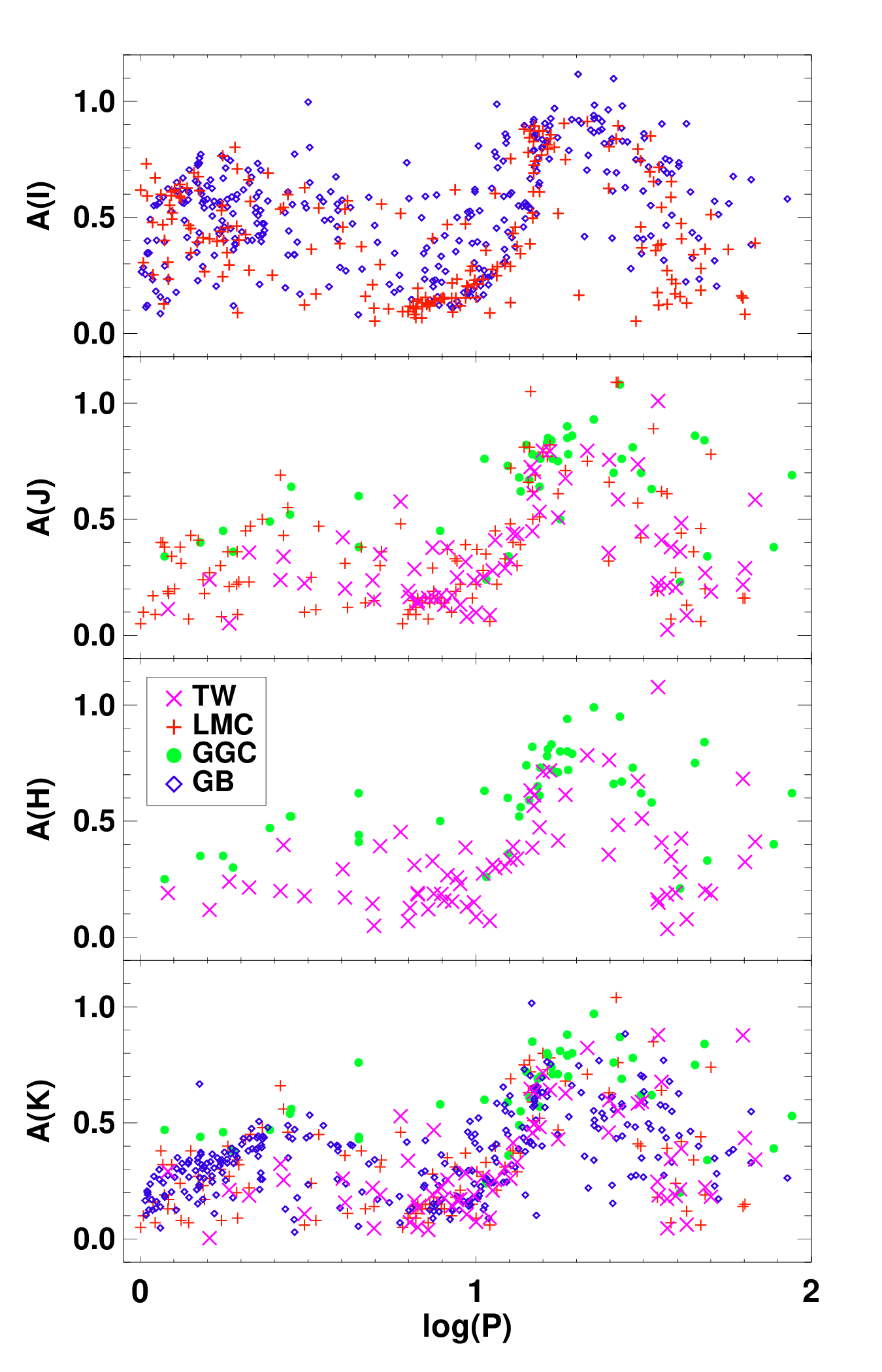}
\caption{ Amplitudes at various wavelengths for Type \rom{2} Cepheids in the Galactic Bulge \citep[GB,][]{soszynski2011, minnitivvv}, Galactic globular clusters \citep[GGC,][]{matsunaga2006} and the LMC \citep{soszynski2008a, ripepi2015}. ``TW'' represents the amplitudes for T2Cs in the LMC based on our observations.}
\label{fig:amp_multi}
\end{figure}

\section{Light Curve Analysis}
\label{sec:fou}

We compiled multi-wavelength data available in the literature in order to study the variation in T2C light curve structure as a function of period and bandpass. The sources used were: optical photometry from OGLE-\rom{3} for 203 objects in the LMC \citep{soszynski2008a} and 357 variables in the Bulge \citep{soszynski2011}, NIR light curves of 46 stars in Galactic globular clusters \citep{matsunaga2006}, and $K_s$-band light curves of 130 variables in the LMC \citep{ripepi2015}. We also cross-matched the OGLE-\rom{3} catalog of Bulge T2Cs against the latest catalog (DR4, Hempel et al.~, in prep.) from the VVV survey \citep{minnitivvv}. We obtained $\sim225$ good quality light curves with an average of 50 epochs in the $K_s$-band, which were used in the analysis presented below. Details regarding the cross-match, selection criteria, photometry and other properties will be presented in a separate study (Bhardwaj et al.~, in prep.). 

We fit each $IJHK_s$ light curve with a fourth-order Fourier sine series, $m = m_0 + \sum_{i=1}^{4}A_{i}\sin(2\pi\phi + \Phi_{i})$ \citep{bhardwaj2015}, where $m$ is the observed magnitude, $A_i$ is the amplitude of each term and $\phi$ represents the corresponding phase. The series is kept to $i\leq 4$ because the number of observations is small and most light curves have large gaps in phase. Fig.~\ref{fig:amp_multi} shows the total amplitude of each variable in each band as a function of period. We also plot the amplitudes derived by \citet{ripepi2015} for comparison. The amplitude increases as a function of period for WVI stars with periods 8-20~d, while it exhibits the opposite behavior for RVT variables. The $I$-band amplitudes are the best determined since those light curves are of much higher quality.  The short-period BLH stars are fainter and the long-period RVT stars exhibit irregular light curves. Therefore, the amplitudes for these variables display a greater scatter as compared to WVI stars. A more detailed discussion on variation of light curve parameters as a function of period and wavelength for (Classical) Cepheids is presented in \citet{bhardwaj2017}.

\subsection{Templates for Type \rom{2} Cepheids}

\begin{figure}[t]
\begin{center}
\includegraphics[width=0.5\textwidth,keepaspectratio]{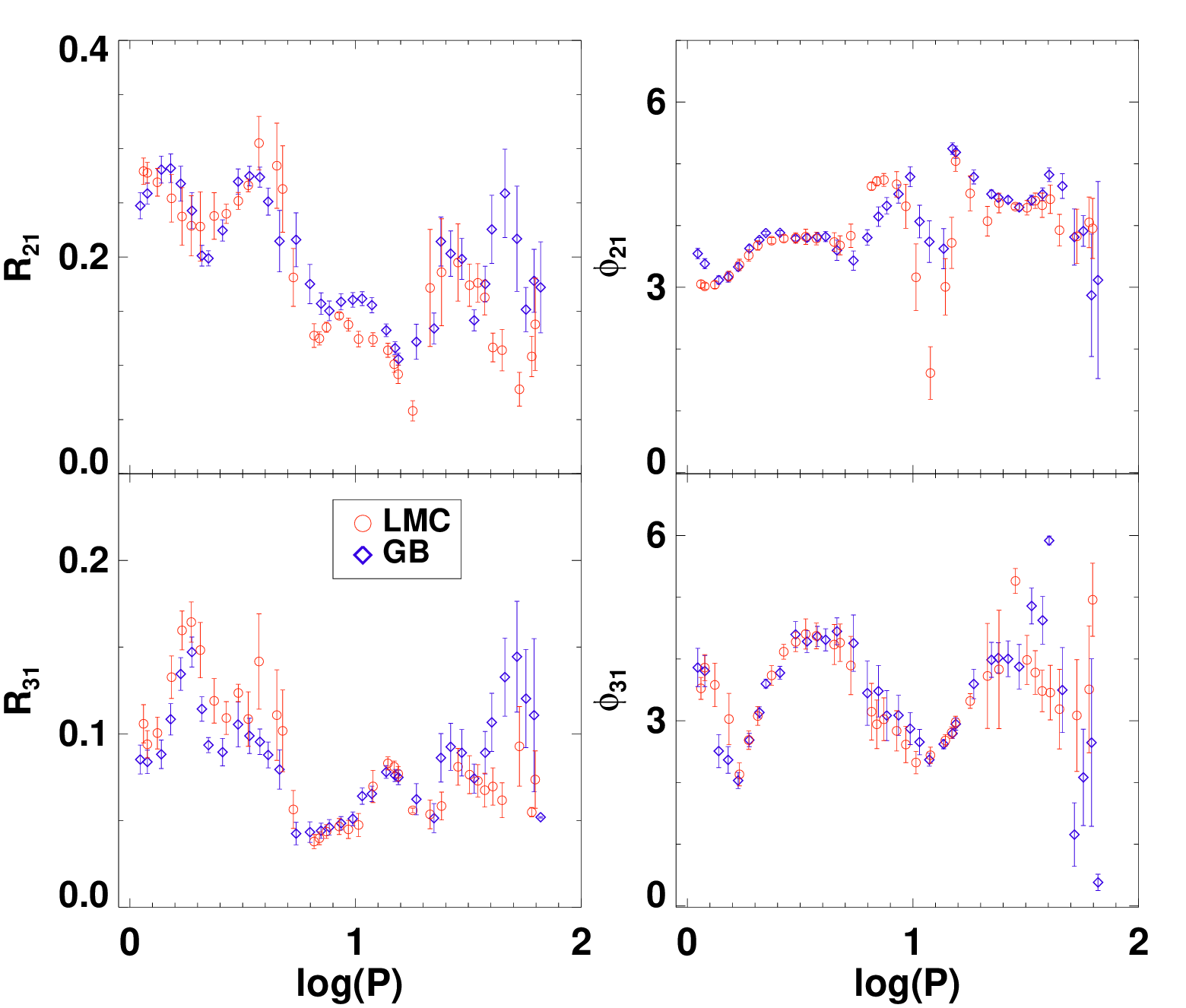}
\caption{Mean Fourier parameters in $I$-band for Type \rom{2} Cepheids in the Galactic Bulge and the LMC. The $\phi_{31}$ parameter is converted to a cosine series for plotting purposes.}
\label{fig:fou_t2c}
\end{center}
\end{figure}

The NIR photometry of T2Cs available in the literature does not have sufficient phase coverage or photometric accuracy to construct light curve templates for these type of variables. For example, the light curves of T2Cs in the LMC currently available from the VMC survey  have an average of 12 epochs in the $K_s$-band, while the $J$-band photometry is limited to only a couple of epochs \citep{ripepi2015}. T2Cs in Galactic globular clusters have 9-40 observations per light curve but the sample is limited to 46 stars \citep{matsunaga2006}. There are no other near-infrared time-series studies on T2Cs in the literature, thus, limiting the sample size and the phase coverage for each period range. Therefore, we use OGLE-\rom{3} LMC $I$-band data for the purpose of constructing templates. We also use $K_s$-band photometry from VVV to construct an alternative set of templates for comparison. Fig.~\ref{fig:fou_t2c} displays the $I$-band Fourier parameters for T2Cs in the LMC and Bulge. Note that the coefficients associated with the lower order Fourier amplitudes ($R_{21}$ and $R_{31}$) and phases ($\phi_{21}$ and $\phi_{31}$) contain most of the quantitative information about light curve structure \citep{slee81, bhardwaj2017}. The mean Fourier parameters exhibit similar variations as a function of period for both LMC and Bulge samples in the $I$-band. Therefore, it is reasonable to assume that the light curve structure in the $K_s$-band is also similar for both populations, as metallicity effects are less pronounced at longer wavelengths. The final calibrator sample consists of 170 $I$-band and 225 $K_s$-band light curves in the LMC and Bulge, respectively. We divide the $I$ and $K_s$-band light curves into 10 period bins as the mean Fourier parameters vary significantly as a function of this parameter. The period is the best observable to trace the changes in light curve structure because it is independent of the wavelength. The adopted period bins and the number of stars in each bin are listed in Table~\ref{table:pbin}. The number of stars per bin ranges from 8 to 69, with a significant fraction consisting of BLH stars having $P\lesssim 2$~d. The median photometric uncertainty per bin is $\sim0.01$ and $\sim0.1$~mag for $I$ and $K_s$-band, respectively.

\begin{deluxetable}{cccc}
\tablewidth{0.5\textwidth}
\tablecaption{Adopted period bins \label{table:pbin}}
\tablehead{\colhead{~~~~~Bin~~~~~} & \colhead {~~~~~$P$~(d)~~~~~} & \colhead {~~~~~$N_I$~~~~~} & \colhead{~~~~~$N_{K_s}$~~~~~}}
\startdata
 1&  1-2   &  48 & 69 \\
 2&  2-3   &  11 & 30 \\
 3&  3-5   &  10 & 18 \\
 4&  5-7   &  15 &  8 \\
 5&  7-9   &  21 & 21 \\
 6&  9-11  &  12 & 22 \\
 7& 11-13  &   9 & 14 \\
 8& 13-15  &  12 & 16 \\
 9& 15-20  &  15 & 13 \\
10& $>$20  &  17 & 14\\
\enddata
\tablecomments{ Period range for each bin is provided in second column. $N_I$ and $N_{K_s}$ represent the number of stars in each bin with good-quality light-curves in $I$ and $K_s$-band, respectively.}
\end{deluxetable}

\begin{figure}[t]
\begin{center}
\includegraphics[width=0.5\textwidth,keepaspectratio]{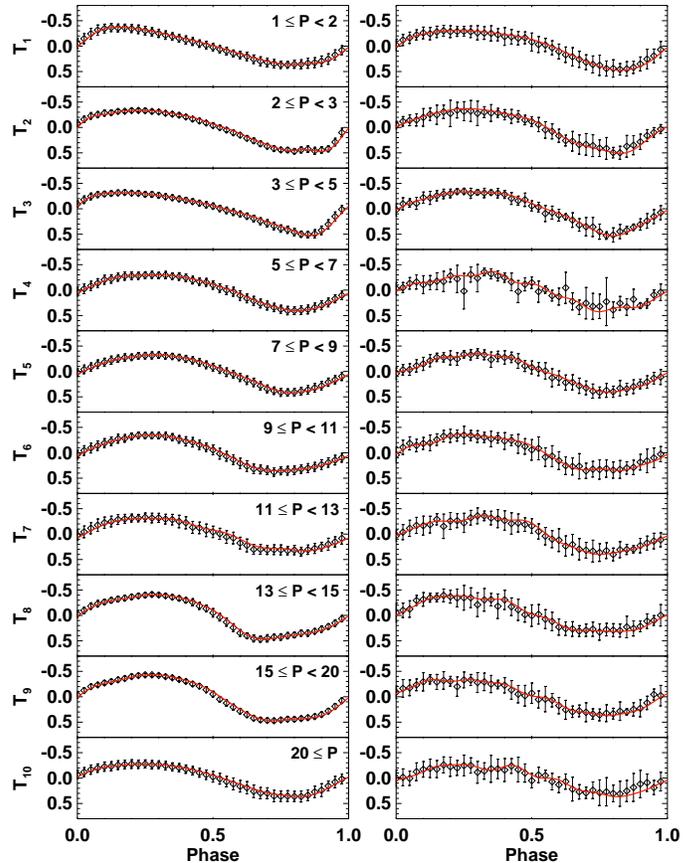}
\caption{Median light curves for different period bins, based on observations of Type \rom{2} Cepheids in the LMC at $I$-band (left) and the Galactic Bulge at $K_s$-band (right). The binning step size is $0.025$ in phase, with the average value and standard deviation of the mean displayed with diamond symbols and error bars. The solid red lines represent seventh-order Fourier fits. The adopted range of periods (in days) for each bin, is labeled in the top right corner of the left panel.}
\label{fig:norm_lc}
\end{center}
\end{figure}

\citet{inno2015} recently derived near-infrared templates for Classical Cepheids and suggested that setting the zero phase of the light curves to the epoch of mean brightness would avoid problems in estimating precise maximum for bump Cepheid light curves with poor phase coverage. Although the calibrating sample of T2Cs has very good phase coverage in both $I$ and $K_s$-band, we adopt the same phasing strategy to avoid any complications. We first fit light curves with a fourth-order Fourier series and determine the phase corresponding to mean magnitude along the rising branch. We normalize the light curves to zero mean and unity amplitude and merge those within an adopted period bin. We then adopt a seventh-order Fourier series as the optimum fit, following previous work on templates by \citet{sosz2005, inno2015}. The residuals from these fits follow a normal distribution and we recursively remove $3\sigma$ outliers to increase the robustness of our results. The merged light curves and the Fourier series fits are displayed in Fig.~\ref{fig:norm_lc}, while the Fourier coefficients are listed in Table~\ref{table:temp_t2c}. $T_1,T_2,.....T_{10}$ represent the merged light curve templates in each bin. Typical standard deviations of the template fits are $\lesssim0.01$~mag and $0.01$ in $I$ and $K_s$-band, respectively. It is evident that the progression of the normalized and merged templates in each bin is similar for both bands, although the $K_s$-band templates are based on a significantly smaller number of data points. While the seventh-order series fits result in some wiggles for the $K_s$-band templates, it has no impact on the derived mean magnitudes. For example, the Fourier fits for the $T_4$ and $T_{10}$ sets could have been done at lower order, but we decided to retain the same expansion to facilitate the comparison with $I$-band templates. 

\section{Leavitt Laws for Type \rom{2} Cepheids}
\label{sec:pl_t2c}

\begin{deluxetable*}{lrrrrrrrrrrrrrrr}
\tablewidth{0pt}
\tablecaption{Fourier coefficients of the light-curve templates for Type \rom{2} Cepheids. \label{table:temp_t2c}}
\tablehead{\colhead{Bin}&\colhead{$A_1$}&\colhead{$A_2$}&\colhead{$A_3$}&\colhead{$A_4$}&\colhead{$A_5$}&\colhead{$A_6$}&\colhead{$A_7$}&\colhead{$\phi_1$}&\colhead{$\phi_2$}&\colhead{ $\phi_3$}&\colhead{ $\phi_4$}&\colhead{ $\phi_5$}&\colhead{ $\phi_6$}&\colhead{ $\phi_7$}&\colhead{ $\sigma$}}
\startdata
\multicolumn{16}{c}{$I$-band}\\
 1& 0.361& 0.088& 0.033& 0.019& 0.010& 0.002& 0.002& 3.138& 3.164& 2.835& 2.822& 2.855& 3.295& 0.045& 0.005\\
 2& 0.408& 0.091& 0.042& 0.030& 0.026& 0.017& 0.011& 2.944& 3.412& 3.429& 3.462& 3.693& 3.914& 4.382& 0.002\\
 3& 0.365& 0.135& 0.061& 0.027& 0.010& 0.002& 0.002& 2.981& 3.426& 3.841& 4.250& 4.589& 5.563& 1.304& 0.003\\
 4& 0.352& 0.052& 0.009& 0.000& 0.004& 0.003& 0.003& 2.879& 3.716& 3.801& 1.192& 5.833& 4.261& 4.756& 0.005\\
 5& 0.362& 0.046& 0.011& 0.006& 0.005& 0.003& 0.001& 2.881& 3.898& 5.045& 1.545& 3.287& 4.830& 6.009& 0.004\\
 6& 0.357& 0.009& 0.016& 0.011& 0.003& 0.003& 0.001& 2.986& 4.778& 1.775& 3.465& 0.874& 2.383& 2.968& 0.005\\
 7& 0.332& 0.035& 0.012& 0.016& 0.013& 0.007& 0.003& 2.928& 3.099& 2.589& 2.903& 5.821& 1.984& 2.012& 0.008\\
 8& 0.446& 0.043& 0.041& 0.027& 0.004& 0.004& 0.001& 3.079& 5.385& 2.518& 3.832& 0.974& 2.593& 6.001& 0.002\\
 9& 0.471& 0.028& 0.030& 0.028& 0.011& 0.001& 0.002& 3.030& 4.583& 2.997& 3.836& 3.793& 3.646& 2.417& 0.001\\
10& 0.318& 0.055& 0.020& 0.006& 0.002& 0.004& 0.002& 3.044& 3.860& 4.047& 4.748& 6.039& 1.181& 2.433& 0.007\\
\tableline
\multicolumn{16}{c}{$K_s$-band}\\
 1& 0.381& 0.100& 0.024& 0.012& 0.005& 0.004& 0.002& 2.895& 3.661& 3.681& 3.801& 3.629& 3.714& 3.394& 0.009\\
 2& 0.421& 0.071& 0.015& 0.025& 0.012& 0.005& 0.002& 2.957& 4.093& 4.509& 4.984& 0.043& 0.622& 1.530& 0.008\\
 3& 0.405& 0.073& 0.028& 0.016& 0.006& 0.006& 0.003& 2.914& 4.032& 4.910& 0.408& 3.104& 2.779& 3.042& 0.005\\
 4& 0.351& 0.043& 0.018& 0.023& 0.012& 0.024& 0.021& 2.895& 4.366& 4.554& 2.869& 4.454& 4.199& 0.961& 0.013\\
 5& 0.363& 0.030& 0.010& 0.010& 0.007& 0.011& 0.013& 2.885& 4.292& 4.832& 0.594& 4.471& 5.421& 4.353& 0.008\\
 6& 0.362& 0.026& 0.017& 0.028& 0.007& 0.013& 0.011& 2.958& 4.837& 2.379& 4.467& 3.568& 2.370& 2.487& 0.013\\
 7& 0.365& 0.060& 0.022& 0.008& 0.018& 0.018& 0.004& 2.897& 4.832& 2.356& 3.440& 1.580& 5.186& 2.610& 0.013\\
 8& 0.371& 0.020& 0.028& 0.004& 0.016& 0.011& 0.012& 3.091& 3.158& 2.789& 0.581& 5.449& 1.873& 5.638& 0.013\\
 9& 0.357& 0.047& 0.017& 0.013& 0.011& 0.007& 0.017& 3.129& 3.903& 3.743& 2.693& 5.985& 1.475& 5.316& 0.012\\
10& 0.297& 0.051& 0.009& 0.023& 0.019& 0.020& 0.017& 2.874& 3.766& 1.938& 6.167& 4.367& 0.966& 4.343& {0.017}\\
\enddata
\end{deluxetable*}

\begin{deluxetable*}{lrlrrrrrrrrc}
\tablewidth{0pt}
\tablecaption{Properties of Type \rom{2} Cepheids in the LMC. \label{table:t2c}}
\tablehead{\colhead{ID}&\colhead{P (d)}&\colhead{Class}&\multicolumn{5}{c}{Mean magnitudes}&\multicolumn{3}{c}{$\sigma$}&\colhead{$E(V-I)$}\\
\colhead{}&\colhead{}&\colhead{}&\colhead{$V$}&\colhead{$I$}&\colhead{$J$}&\colhead{$H$}&\colhead{$K_s$}&\colhead{$J$}&\colhead{$H$}&\colhead{$K_s$}&\colhead{[mag]}}
\startdata
OGLE-LMC-T2CEP-025& 67.965& RVT& 15.102& 14.042& 13.554& 13.209& 13.160& 0.059& 0.084& 0.120& 0.070\\
OGLE-LMC-T2CEP-028&  8.785& PWV& 16.045& 15.543& 15.340& 14.881& 14.732& 0.056& 0.045& 0.072& 0.050\\
OGLE-LMC-T2CEP-029& 31.245& RVT& 15.446& 14.642& 14.020& 13.607& 13.354& 0.028& 0.040& 0.041& 0.040\\
OGLE-LMC-T2CEP-033&  9.395& PWV& 16.468& 15.788& 15.253& 14.887& 14.845& 0.044& 0.055& 0.054& 0.050\\
OGLE-LMC-T2CEP-034& 14.911& WVI& 17.317& 16.228& 15.485& 14.971& 14.797& 0.057& 0.054& 0.048& 0.080\\
OGLE-LMC-T2CEP-035&  9.866& WVI& 17.162& 16.259& 15.637& 15.127& 15.048& 0.050& 0.054& 0.078& 0.050\\
OGLE-LMC-T2CEP-036& 14.881& WVI& 16.745& 15.845& 15.161& 14.692& 14.634& 0.036& 0.053& 0.059& 0.050\\
\enddata
\tablecomments{Star ID, period, class and mean $VI$ magnitudes are taken from OGLE-\rom{3} \citet{soszynski2008a}. $E(V-I)$ values are based on the maps of \citet{hasch11}. The entire table is available online as supplemental material; only a few lines are shown here for guidance regarding its content.}
\end{deluxetable*}

\begin{figure*}
\begin{center}
\includegraphics[width=1.0\textwidth,keepaspectratio]{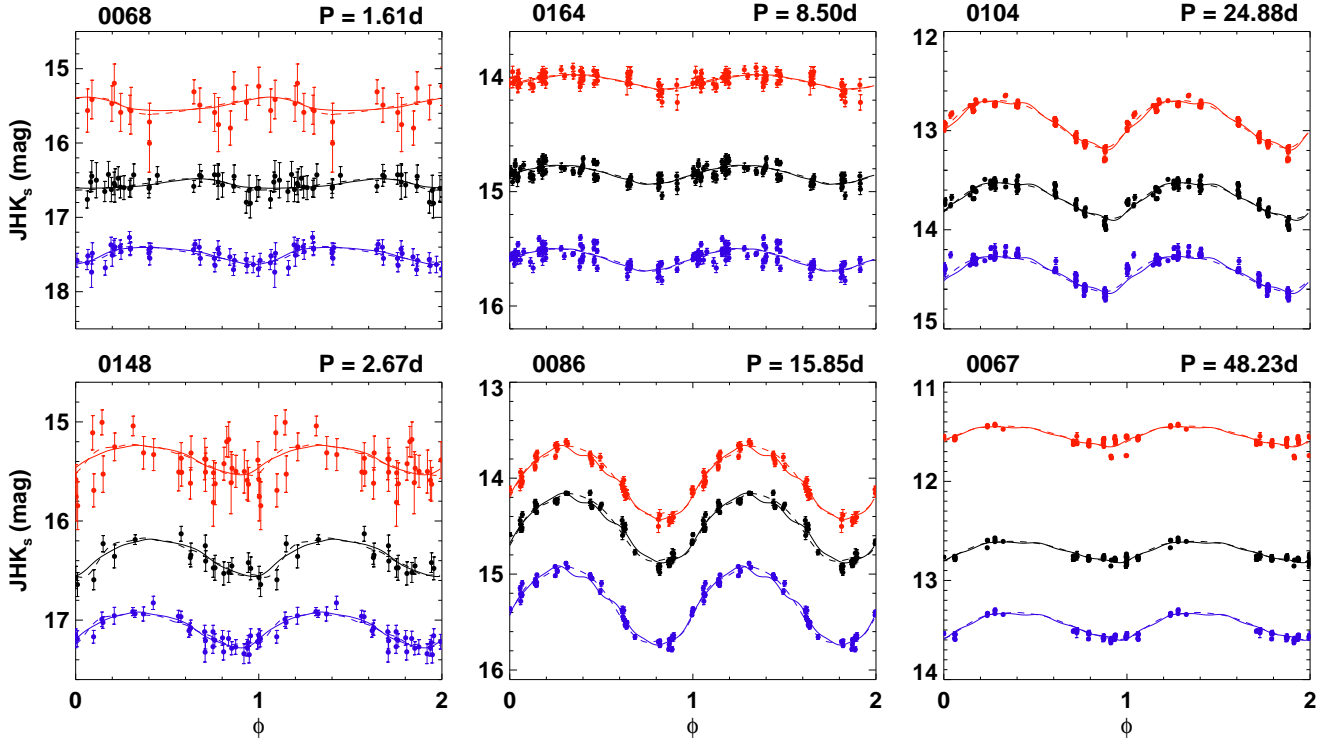}
\caption{Representative near-infrared light curves of six Type \rom{2} Cepheids in the LMC. The left, middle and right panels show short, intermediate and long-period BLH, WVI and RVT stars, respectively. $J$, $H$ and $K_s$-band light curves are plotted using blue, black and red symbols. The $J$ and $K_s$-band light curves are offset by +0.25 and -0.5 mag, respectively. The solid and dashed lines represent the $I$ and $K_s$-band templates, respectively, fit to the data in each band.}
\label{fig:lc_nir}
\end{center}
\end{figure*}

\begin{figure*}[t]
\begin{center}
\includegraphics[height=8in,keepaspectratio]{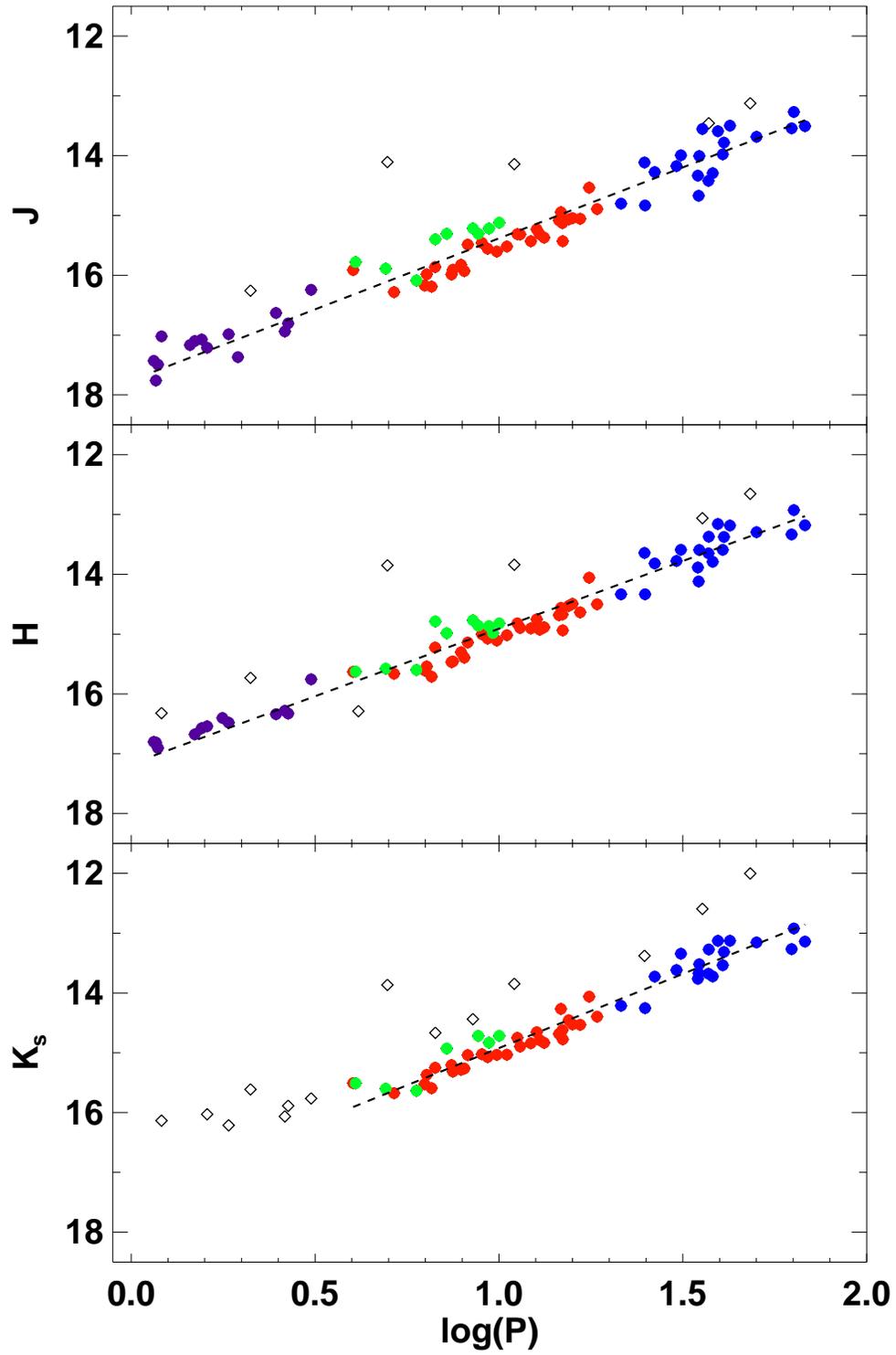}
\caption{Extinction-corrected near-infrared P-L relations for Type \rom{2} Cepheids in the LMC, based exclusively on our photometry. The violet, red, green and blue symbols represent BLH, WVI, PWV and RVT stars, respectively. The dashed line represents a single regression line over the entire period range and open diamonds show $2\sigma$ outliers.}
\label{fig:pl_nir}
\end{center}
\end{figure*}

\begin{deluxetable*}{llrrrrrr}
\tablewidth{6.5in}
\tablecaption{Type \rom{2} Cepheid P-L Relations in the LMC based exclusively on our photometry\label{table:lmc_pl}.}
\tablehead{\colhead{Band} & \colhead{Types}& \colhead{$a$} & \colhead{$\sigma_a$} & \colhead{$b$} & \colhead{$\sigma_b$} &\colhead{\it rms} & \colhead{$N$}}
\startdata
$J$  & B+W& -2.100 & 0.107 & 15.539 & 0.038 & 0.149& 40\\
$J$  & P+R& -2.256 & 0.111 & 15.168 & 0.059 & 0.274& 30\\
$J$  & all& -2.374 & 0.058 & 15.383 & 0.026 & 0.259& 73\\
$H$  & B+W& -1.963 & 0.109 & 15.053 & 0.042 & 0.110& 39\\
$H$  & P+R& -2.188 & 0.116 & 14.767 & 0.060 & 0.239& 31\\
$H$  & all& -2.261 & 0.061 & 14.908 & 0.028 & 0.208& 72\\
$K_s$& B+W& -2.117 & 0.318 & 14.992 & 0.047 & 0.081& 26\\
$K_s$& P+R& -2.062 & 0.118 & 14.627 & 0.063 & 0.244& 29\\
$K_s$& all& -2.483 & 0.089 & 14.922 & 0.038 & 0.190& {56}\\
\enddata
\tablecomments{B+W: BLH+WVI; P+R: PWI+RVT.}
\end{deluxetable*}

\begin{deluxetable*}{lrrrrllrr}
\tablewidth{6.5in}
\tablecaption{Comparison of Type \rom{2} Cepheid P-L relations.\label{table:comp_pl}}
\tablehead{\colhead{Band} & \colhead{$a$} & \colhead{$\sigma_a$} & \colhead{\it rms} & \colhead{$N$}& \colhead{Loc} & \colhead{Src}& \colhead{$|T|$} & \colhead{$p(t)$}}
\startdata
$J$   & -2.374 & 0.058 & 0.259&  73 & LMC & TW  &  \nd  & \nd  \\
$J$   & -2.163 & 0.044 & 0.210& 137 & LMC & M09 & 3.017 & 0.003\\
$J$   & -2.190 & 0.040 & 0.130& 120 & LMC & R14 & 2.557 & 0.011\\
$J$   & -2.092 & 0.116 & 0.330&  47 & SMC & M11 & 2.344 & 0.021\\
$J$   & -2.230 & 0.053 & 0.160&  46 & GGC & M06 & 1.591 & 0.114\\
$H$   & -2.261 & 0.061 & 0.208&  72 & LMC & TW  &  \nd  & \nd  \\
$H$   & -2.316 & 0.043 & 0.200& 136 & LMC & M09 & 0.746 & 0.457\\
$H$   & -2.214 & 0.148 & 0.320&  25 & SMC & M11 & 0.357 & 0.722\\
$H$   & -2.344 & 0.050 & 0.150&  46 & GGC & M06 & 0.996 & 0.321\\
$K_s$ & -2.483 & 0.089 & 0.190&  56 & LMC & TW  &  \nd  & \nd  \\
$K_s$ & -2.278 & 0.047 & 0.210& 129 & LMC & M09 & 1.933 & 0.055\\
$K_s$ & -2.385 & 0.030 & 0.090& 120 & LMC & R14 & 1.312 & 0.191\\
$K_s$ & -2.113 & 0.105 & 0.290&  45 & SMC & M11 & 2.609 & 0.011\\
$K_s$ & -2.408 & 0.047 & 0.140&  46 & GGC & M06 & 0.768 & 0.444\\
$K_s$ & -2.240 & 0.140 & 0.410&  39 & GB  & G08 & 1.399 & \mcr{0.165}\\
\enddata
\tablecomments{Loc: LMC/SMC: Large/Small Magellanic Cloud, GGC: Galactic globular clusters; GB: Galactic Bulge. Src: TW: this work; M06: \citet{matsunaga2006}; G08: \citet{gmat2008}; M09: \citet{matsunaga2009}; M11: \citet{matsunaga2011}; R15: \citet{ripepi2015}. $|T|$ represents the observed value of the t-statistic and $p(t)$ gives the probability of acceptance of the null hypothesis (equal slopes).}
\end{deluxetable*}

We phase the NIR light curves of T2Cs in the LMC using the OGLE-\rom{3} values of $P$ and $T_{I,\rm {max}}$ \citep{soszynski2008a}. We use the same technique described above to set the zero phase of the light curves to the time of mean light in the rising branch. We fit the templates from Table~\ref{table:temp_t2c} and solve independently for each amplitude and a possible phase shift in the time of mean light relative to $I$-band. The amplitudes derived through this procedure show similar trends to those obtained via Fourier fit and no significant phase shifts are seen. Fig.~\ref{fig:lc_nir} shows representative light curves for each T2C class. Template-fit light curves for all variables in our sample are presented in Fig.~\ref{fig:temp_lc_lmc}. We note that short-period (fainter) objects exhibit larger scatter than long-period (brighter) stars, as expected from photon statistics.

The mean magnitudes are estimated via intensity-weighted integration of the best-fit $I$ and $K_s$-band based templates. The standard errors on the mean magnitudes are based on the {\it rms} of the fits. The difference in the values obtained from $I$ or $K_s$-band templates is $\lesssim 0.01$~mag for $P>8$~d and $\sim0.02$~mag for shorter periods. In the case of a few BLH stars, the difference in $K_s$-band exceeds $0.05$~mag but since the crowding corrections for most of these objects exceeds $0.2$~mag, they will not be used in the final PLR fits. We take the average of both template-fit mean magnitudes as the final value. Table~\ref{table:t2c} lists the T2C  mean magnitudes and uncertainties in each band. The $JHK_s$ mean magnitudes were corrected for extinction using the reddening law of \citet{card89} with $R_V=3.23$ and individual reddening values from the map of \citet{hasch11}. The total-to-selective absorption ratios per unit of $E(V-I)$ are $R_J = 0.69$, $R_H = 0.43$ and $R_K = 0.28$ \citep{bhardwaj2016a}.

We fit PLRs of the following form:
\begin{equation}\label{eq:plr}
m_{\lambda} = a_{\lambda}[\log(P) - 1] +b_{\lambda},
\end{equation}
\noindent where $m_\lambda$ is the extinction-corrected mean magnitude, $\lambda$ represents the $JHK_s$ bands, $a$ is the slope and $b$ is the zeropoint at $P=10$~d. We fit PLRs to the entire sample as well as to subsamples of faint (BLH+WVI) and bright (PWV+RVT) variables, iteratively removing $2\sigma$ outliers in all cases. As most of the outliers are likely due to blends or additional crowding effects, they appear on the bright side of the PLRs \citep[see also discussion in][]{matsunaga2009, ripepi2015}. We adopt this threshold throughout the paper to have a stronger constraint on slopes and zeropoints. Since the samples are small, a higher sigma-clipping threshold marginally changes the slopes (by less than the half of their quoted uncertainties) and the typical increase in the number of stars and the dispersion is less than $10\%$. Fig.~\ref{fig:pl_nir} displays the results of the fits, which are also summarized in Table~\ref{table:lmc_pl}. We also note that a detailed statistical analysis on P-L relations was presented in Paper III to test Classical Cepheid data for non-linearity under various assumptions such as independent identically distributed observations, normality of residuals and homoskedasticity. We also performed White's test \citep{white1980} for Type II Cepheid sample and found that our data provide evidence of homoskedasticity under 95\% confidence interval.

Previous studies at optical and NIR wavelengths \citep[see,][]{soszynski2008a, matsunaga2009, ripepi2015} have suggested that PWV and RVT stars lie above the PLR defined by the shorter-period BLH and WVI stars. A single PLR fit to our entire sample also gives evidence that WVI stars are mostly found below the regression line, specially in $J$-band. We use the $F$-test as described in Paper III, to quantify the statistical significance of non-linearities in the slopes of the PLRs for various subsamples. We find a considerable difference between the PLR slopes for RVTs and BLH+WVI variables when considered separately. However, we find consistent slopes between PWV+RVT stars and BLH+WVI variables. Table~\ref{table:lmc_pl} summarizes our findings. We note that the PLR dispersions are reduced by $12\%$, $2\%$ and $3\%$ in $JHK_s$, respectively, when using template-fit instead of Fourier-fit magnitudes.

\subsection{Comparison with published P-L relations}

We compare our PLRs to previous work carried out by \citet{matsunaga2006},  \citet{gmat2008}, \citet{matsunaga2009},  \citet{matsunaga2011}, and \citet{ripepi2015}. We use the t-test as discussed in Paper II, to compare the slopes given their uncertainties and the dispersion of the underlying relations. The observed t-statistic is compared with theoretical values, calculated from the t-distribution at the 95\% confidence interval (see Paper \rom{2} for details). In brief, the null hypothesis that the two slopes are the same is rejected if the observed t-statistic ($|T|$) is greater than the theoretical (t) value. Table~\ref{table:comp_pl} lists the various slopes and the results of the test-statistic. The probability $(p(t))$ of the acceptance of null hypothesis is also provided and $p(t)\!<\!0.05$ suggests that the two slopes under consideration are not equal.

We find that the slope of the $J$-band PLR for our entire sample is not consistent with those derived by \citet{matsunaga2009, ripepi2015} for T2Cs in the LMC but it is consistent in $H$ and $K_s$-bands. We note that those studies did not consider RVTs since they were found to lie well above the single regression line relative to shorter-period T2Cs at optical wavelengths. This deviation is not significant in our sample in $K_s$-band. The slope of the $J$-band PLR for BLH+WVI stars is $-2.100\pm0.107$, consistent with published work given the uncertainties. The slopes of the LMC PLRs are not consistent with their SMC counterparts in $J$ and $K_s$-bands from \citet{matsunaga2011} but are in agreement in all bands with those from globular clusters \citep{matsunaga2009}. Furthermore, the slope of the LMC $K_s$-band PLR is also consistent with the corresponding relation based on Bulge variables \citep{gmat2008}.

\begin{figure}
\begin{center}
\includegraphics[width=0.5\textwidth,keepaspectratio]{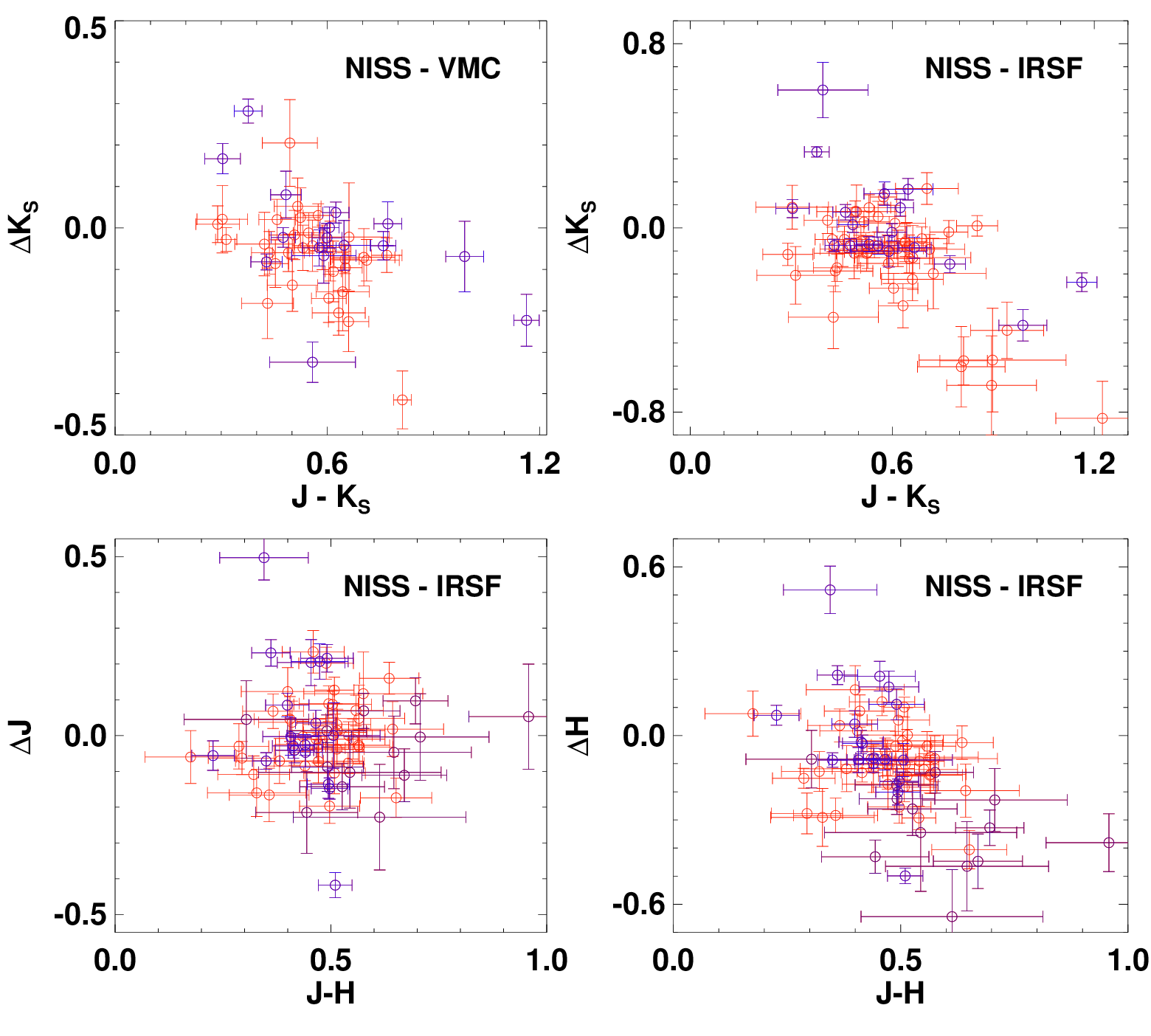}
\caption{{\it Top left:} Comparison of $K_s$-band mean magnitudes as a function of $J-K_s$ for stars in common with \citet{ripepi2015}. {\it Top right:} Same as top-left panel but with random-phase corrected mean magnitudes from \citet{matsunaga2009}. {\it Bottom panels:} Same as top-right but for $J$ and $H$-band magnitudes as a function of $J-H$.}
\label{fig:com_mag}
\end{center}
\end{figure}

We also compare our T2C NIR mean magnitudes with values found in the literature. We find 76 objects in common with \citet{matsunaga2011} and 62 with \citep{ripepi2015}. We note that the former are single-epoch $JHK_s$ measurements in the IRSF \citep{kato2007} system while the latter are mean magnitudes in the VISTA system. We apply the relevant color transformations from IRSF to 2MASS following \citep{kato2007} and from VISTA to 2MASS as derived by \citet{CASU}. Fig.~\ref{fig:com_mag} shows the difference in our magnitudes with respect to VISTA and IRSF as a function of color. The $K_s$-band mean magnitudes in this work are consistent with those from VISTA given their uncertainties. The agreement with IRSF is also good except for a few stars with $J-K_s>0.7$~mag. The $J$-band magnitudes from our work and IRSF are in good agreement, while a mild trend is seen in $H$-band.

\subsection{Near-infrared P-L and P-W relations for OGLE-\rom{3} sample of Type \rom{2} Cepheids \label{ssec:pl_mxd}}

\begin{figure*}[t]
\begin{center}
\includegraphics[width=0.63\textwidth,keepaspectratio]{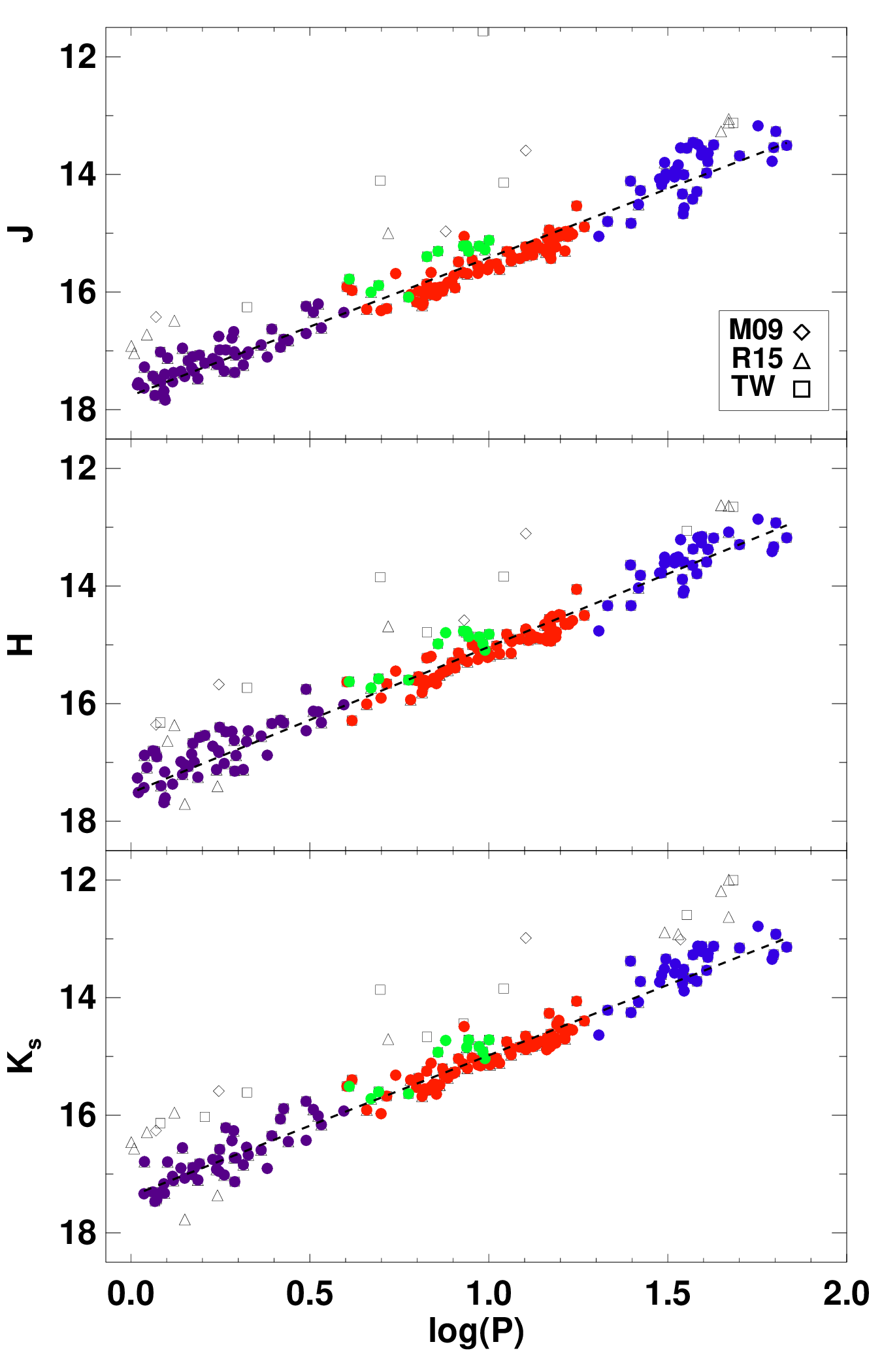}
\caption{Extinction-corrected near-infrared P-L relations for OGLE-\rom{3} Type \rom{2} Cepheids in the LMC with photometry from our work and the literature. The violet, red, green and blue colors represent BLH, WVI, PWV and RVT stars, respectively. The square, triangle, and diamond symbols represent the photometry from this work (TW), \citet[R15,][]{ripepi2015} and \citet[M09,][]{matsunaga2009}, respectively. The dashed line represents a single regression line over the entire period range and empty symbols show 2$\sigma$ outliers.}
\label{fig:pl_mxd}
\end{center}
\end{figure*}

\begin{deluxetable*}{llrrrrrr}
\tablewidth{8in}
\tabletypesize{\footnotesize}
\tablecaption{Near-infrared P-L and P-W relations for Type \rom{2} Cepheids in the LMC, based on mixed photometry. \label{table:pl_mxd}}
\tablehead{\colhead{Band} & \colhead{Type} & \colhead{$a$} & \colhead{$\sigma_a$}  & \colhead{$b$} & \colhead{$\sigma_b$} & \colhead{\it rms} & \colhead{$N$}}
\startdata
$J$        & BLH & -2.294 & 0.153 & 15.375 & 0.113 & 0.202&  55\\
           & WVI & -2.378 & 0.105 & 15.580 & 0.018 & 0.111&  72\\
           & B+W & -2.061 & 0.038 & 15.563 & 0.017 & 0.157& 126\\
           & P+R & -2.249 & 0.072 & 15.144 & 0.039 & 0.320&  53\\
           & all & -2.346 & 0.025 & 15.416 & 0.012 & 0.252& 180\\
$H$        & BLH & -2.088 & 0.214 & 15.218 & 0.163 & 0.296&  52\\
           & WVI & -2.457 & 0.111 & 15.160 & 0.018 & 0.123&  72\\
           & B+W & -2.202 & 0.046 & 15.142 & 0.017 & 0.171& 117\\
           & P+R & -2.297 & 0.071 & 14.752 & 0.038 & 0.269&  52\\
           & all & -2.484 & 0.029 & 15.034 & 0.013 & 0.241& 174\\
$K_s$      & BLH & -2.083 & 0.154 & 15.162 & 0.114 & 0.262&  47\\
           & WVI & -2.250 & 0.097 & 15.078 & 0.016 & 0.119&  72\\
           & B+W & -2.232 & 0.037 & 15.070 & 0.015 & 0.180& 119\\
           & P+R & -2.173 & 0.071 & 14.654 & 0.036 & 0.309&  51\\
           & all & -2.395 & 0.027 & 14.981 & 0.012 & 0.228& 167\\
$W_{J,H}$  & B+W & -2.437 & 0.071 & 14.455 & 0.029 & 0.328& 119\\
           & P+R & -2.174 & 0.112 & 14.095 & 0.061 & 0.337&  53\\
           & all & -2.548 & 0.046 & 14.391 & 0.021 & 0.338& 172\\
$W_{J,K_s}$& B+W & -2.346 & 0.051 & 14.724 & 0.021 & 0.216& 119\\
           & P+R & -2.216 & 0.088 & 14.349 & 0.046 & 0.345&  50\\
           & all & -2.529 & 0.034 & 14.648 & 0.015 & 0.249& 166\\
$W_{H,K_s}$& B+W & -2.248 & 0.085 & 14.959 & 0.030 & 0.321& 115\\
           & P+R & -2.155 & 0.127 & 14.576 & 0.064 & 0.304&  45\\
           & all & -2.478 & 0.057 & 14.832 & 0.023 & 0.369& 165\\
$W_{V,J}$  & B+W & -2.269 & 0.038 & 14.957 & 0.018 & 0.207& 130\\
           & P+R & -2.292 & 0.074 & 14.593 & 0.040 & 0.303&  53\\
           & all & -2.486 & 0.025 & 14.845 & 0.012 & 0.264& 184\\
$W_{V,H}$  & B+W & -2.328 & 0.044 & 14.724 & 0.017 & 0.221& 125\\
           & P+R & -2.388 & 0.072 & 14.442 & 0.039 & 0.268&  52\\
           & all & -2.547 & 0.028 & 14.620 & 0.013 & 0.266& 180\\
$W_{V,K_s}$& B+W & -2.281 & 0.036 & 14.803 & 0.016 & 0.245& 124\\
           & P+R & -2.162 & 0.071 & 14.407 & 0.036 & 0.302&  50\\
           & all & -2.456 & 0.025 & 14.689 & 0.012 & 0.308& 179\\
$W_{I,J}$  & B+W & -2.267 & 0.045 & 15.003 & 0.020 & 0.189& 123\\
           & P+R & -2.277 & 0.083 & 14.588 & 0.045 & 0.315&  52\\
           & all & -2.474 & 0.029 & 14.866 & 0.014 & 0.287& 182\\
$W_{I,H}$  & B+W & -2.351 & 0.048 & 14.709 & 0.018 & 0.215& 119\\
           & P+R & -2.228 & 0.070 & 14.356 & 0.037 & 0.289&  52\\
           & all & -2.503 & 0.031 & 14.631 & 0.013 & 0.239& 169\\
$W_{I,K_s}$& B+W & -2.342 & 0.039 & 14.797 & 0.016 & 0.180& 115\\
           & P+R & -2.148 & 0.071 & 14.392 & 0.036 & 0.312&  50\\
           & all & -2.486 & 0.027 & 14.726 & 0.012 & 0.244&\mcr{167}\\
\enddata
\tablecomments{B+W: BLH+WVI; P+R: PWI+RVT.}
\end{deluxetable*}

\begin{figure*}[t]
\begin{center}
\includegraphics[width=1.0\textwidth,keepaspectratio]{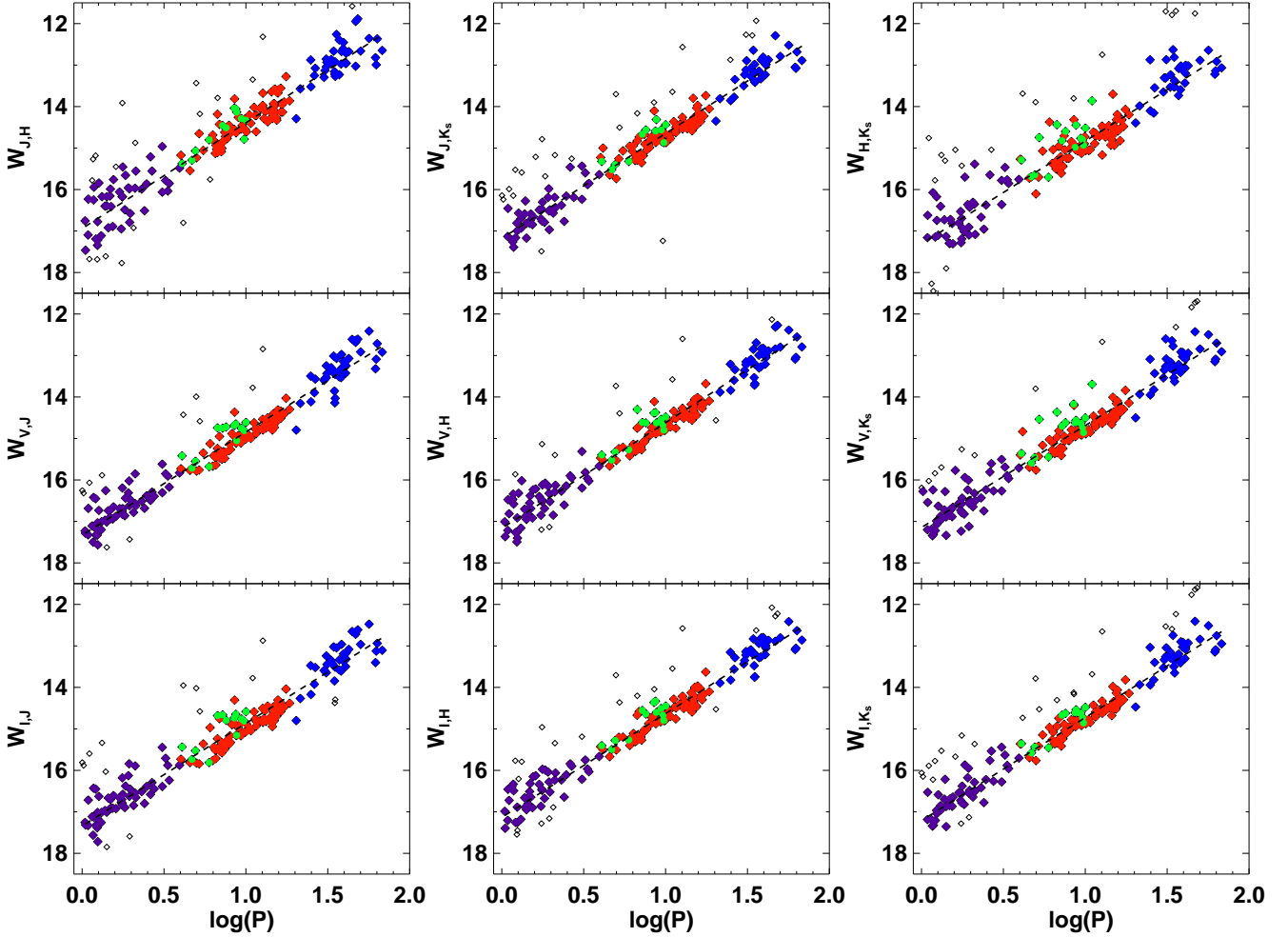}
\caption{Near-infrared Period-Wesenheit relations for Type \rom{2} Cepheids in the LMC. The violet, red, green and blue symbols represent BLH, WVI, PWV and RVT stars, respectively. The dashed line represents a single regression line over the entire period range and open diamonds show $2\sigma$ outliers.}
\label{fig:pw_nir}
\end{center}
\end{figure*}

\begin{deluxetable*}{llrrrrrrrl}
\tablewidth{10in}
\tablecaption{Type \rom{2} Cepheids and RR Lyrae with parallaxes and pulsation distances. \label{table:pi}}
\tablehead{\colhead{ID} & \colhead{Type} & \colhead{$\log P$} & \colhead{$K_s$} & \colhead{$E_{BV}$} & \colhead{$\pi$} & \colhead {\it LKH} & \colhead{[Fe/H]} & \colhead{$\sigma_{\mathrm{[Fe/H]}}$}& \colhead{Src}\\
\colhead{} & \colhead{} & \colhead{[d]} & \multicolumn{2}{c}{[mag]} & \colhead{[mas]} & \colhead{[mag]} & \multicolumn{2}{c}{[dex]} &\colhead{}
}
\startdata
VY   Pyx&  BLH  &  0.093& 5.72& 0.05 & 3.85$\pm$0.28& -0.01& -0.01 &  0.15& {\it Gaia}\\
SW   Tau&  BLH  &  0.200& 7.95& 0.28 & 1.37$\pm$0.04&  \nd &  0.22 &  \nd & B-W \\
V553 Cen&  BLH  &  0.314& 6.86& 0.00 & 1.85$\pm$0.05&  \nd &  0.24 &  \nd & B-W  \\
$k$  Pav&  WVI  &  0.958& 2.78& 0.02 & 5.57$\pm$0.28& -0.02&  0.00 &  0.13& {\it HST} \\
XZ   Cyg& RRab  & -0.331& 8.72& 0.10 & 1.67$\pm$0.17& -0.09& -1.44 &  0.20& {\it HST} \\
UV   Oct& RRab  & -0.266& 8.30& 0.09 & 1.71$\pm$0.10& -0.03& -1.74 &  0.11& {\it HST} \\
RR   Lyr& RRab  & -0.247& 6.49& 0.03 & 3.77$\pm$0.13& -0.02& -1.39 &  0.13& {\it HST} \\
SU   Dra& RRab  & -0.180& 8.62& 0.01 & 1.42$\pm$0.16& -0.11& -1.80 &  0.20& {\it HST} \\
RZ   Cep&  RRc  & -0.511& 7.88& 0.08 & 2.54$\pm$0.19& -0.05& -1.77 &  0.20& {\it HST}\\
\enddata
\tablecomments{$E_{BV}=E(B-V)$. B-W distances from \citet{feast2008} are converted to parallaxes for relative comparison.}
\end{deluxetable*}

We compile $JHK_s$ magnitudes for T2Cs in the LMC that also have $VI$ mean magnitudes from OGLE-\rom{3}. We give preference to our NIR measurements, except for BLH stars in $K_s$-band. If measurements are not available in our database, we use $JK_s$ mean magnitudes from \citet{ripepi2015} or phase-corrected single-epoch magnitudes from \citet{matsunaga2011} as the lowest-priority source. We thus obtain NIR magnitudes in at least one band for 197 out of the 203 OGLE T2Cs in the LMC. All measurements are transformed into the 2MASS system. We derive PLRs for each class of variable following Eqn.~\ref{eq:plr} as well as relations based on ``Wesenheit'' \citep{madore82} magnitudes:
\begin{eqnarray}
\label{eq:pw_all}
W_{\lambda_{2},\lambda_{1}} & = & m_{\lambda_{1}} - R^{\lambda_{2}}_{\lambda_1} (m_{\lambda_{2}}-m_{\lambda_{1}}), \\
R^{\lambda_{2}}_{\lambda_1} & = & \left[\frac{A_{\lambda_{1}}}{E(m_{\lambda_{2}}-m_{\lambda_{1}})}\right] \nonumber
\end{eqnarray}

\noindent where $m_{\lambda}$ is the mean magnitude in one of $VIJHK_s$ and $\lambda_{1}>\lambda_{2}$. We use the \citet{card89} reddening law and assume a value of total-to-selective absorption of $R_V=3.23$. The resulting absorption ratios in other bands are: $R^{J}_{H}=1.63$, $R^{J}_{K_s}=0.69$, $R^{H}_{K_s}=1.92$, $R^{V}_{J}=0.41$, $R^{V}_{H}=0.22$, $R^{V}_{K_s}=0.13$, $R^{I}_{J}=0.92$, $R^{I}_{H}=0.42$, $R^{I}_{K_s}=0.24$ (see Paper \rom{2}). The Wesenheit magnitudes are fit with $\log(P)$ as an independent variable. The resulting PLRs and PWRs are plotted in Figs.~\ref{fig:pl_mxd} and \ref{fig:pw_nir}, respectively, and summarized in Table~\ref{table:pl_mxd}.

The PWRs formed as a combination of $I$-band and one of $J$, $H$ or $K_s$-band exhibit a smaller dispersion than other NIR relations. Similarly, $W_{J,K_s}$ displays a significantly smaller dispersion than $W_{J,H}$ and $W_{H,K_s}$, presumably due to the dominant sample of template-fit mean magnitudes in $J$ and $K_s$-band instead of the phase-corrected single-epoch magnitudes in $H$-band. We compare the PWR slopes with those of \citet{ripepi2015} and find a consistent value for $W_{V,K_s}$, while our value for $W_{V,J}$ is marginally steeper. However, we note that the \citet{ripepi2015} results are based on a single regression fit to BLH+WVI variables and those are consistent with our results for the same subsample.

\subsection{Distance to the LMC using {\it HST} parallaxes\label{ssec:lmc_mu}}

\begin{figure}[h]
\begin{center}
\includegraphics[width=0.52\textwidth,keepaspectratio]{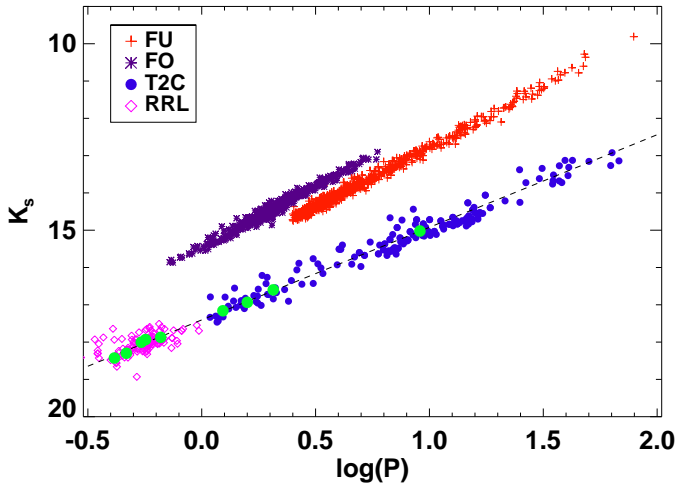}
\caption{A comparison of $K_s$-band P-L relation with Classical Cepheids from LMCNISS data and RR Lyraes in the LMC. The green circles represent the calibrator T2Cs and RRLs.}
\label{fig:t2_rrl}
\end{center}
\end{figure}

\begin{deluxetable}{lr}
\tablewidth{5in}
\tablecaption{Estimates of the LMC distance modulus\label{table:results_mu}}
\tablehead{\colhead{Source}~~ & \colhead{~~$\mu$ [mag]}}
\startdata
T2C $\pi$ & $18.54\pm0.11$\\
RRL $\pi$ & $18.55\pm0.10$\\
T2C B-W   & \mcr{$18.41\pm0.09$}\\
\enddata
\end{deluxetable}

The short-period T2Cs reside in the same instability strip that extends a few magnitudes above the horizontal branch and includes RR Lyraes \citep[RRLs;][]{sandage2006}. It has been suggested that RRLs follow the same PLRs as short period T2Cs \citep{sollima2006, feast2008, ripepi2015}. We therefore further extend the expanded PLRs of \S\ref{ssec:pl_mxd} with NIR measurements of RRLs in the LMC from \citet{borissova2009,muraveva2015}, as shown in Fig.~\ref{fig:t2_rrl}. It can be seen that RRLs nicely follow the PLR of T2Cs, which is shallower than those obeyed by Classical Cepheids.

We use trigonometric parallaxes for 2 T2Cs and 5 RRLs in the solar neighborhood, obtained with the {\it Hubble Space Telescope} \citep[{{\it HST,}\normalfont}][]{benedict2011} and {\it Gaia} \citep{lindegren2016}, to calibrate the zeropoint of our PLRs and estimate a distance to the LMC.  We also use distance estimates for two T2Cs (V553 Cen and SW Tau) determined via the Baade-Wesselink (B-W) method \citep{feast2008}. Table~\ref{table:pi} summarizes the distance estimates for all calibrators along with magnitudes and other properties adopted from \citet{feast2008, benedict2011}. We include the {\it LKH} correction \citep{lutz1973} and ``fundamentalize'' the period of first-overtone RRL by adding $\Delta \log(P) = 0.127$. Lastly, RRL magnitudes are corrected for metallicity effects using the recent $P-L_{K_s}-[{\rm Fe}/\rm{H}]$ relation of \citet{muraveva2015}.

$k$ Pav was classified as a PWV by \citet{feast2008}, but \citet{benedict2011} showed that it follows the same PLR as RRLs. The {\it HST} and {\it Hipparcos} parallaxes of VY Pyx are very dissimilar  \citep[$6.44\pm0.23$~mas and $5.01\pm0.44$~mas from][respectively]{benedict2011,van2007} and yield absolute magnitudes well below the RRL PLR regression line. In contrast, the recent {\it Gaia} parallax \citep[$3.85\pm0.28$~mas,][]{lindegren2016} is in excellent agreement with expectations and is therefore used in our analysis. We note that \citet{benedict2011} provided two parallaxes for the RRL RZ Cep; we adopt $\pi=2.54\pm0.19$~mas since the other choice makes it an outlier in the PLR.

We fix the slope of the $K_s$-band PLR to the value derived from our LMC photometry (``all'' in Table~\ref{table:lmc_pl}) and solve for the difference between the intercept of the Galactic and LMC relations which have absolute and apparent magnitudes, respectively. We thus obtain three estimates of the distance modulus of the LMC based the weighted averages of: (1) the parallaxes of two T2Cs, (2) the parallaxes of five RRLs, and the B-W distances of two T2Cs. The results, listed in Table~\ref{table:results_mu}, are in good agreement with the estimates based on late-type eclipsing binaries \citep[$18.493\pm0.048$~mag,][]{piet13}. These values are also consistent with recent studies \citep{monson12, delmc14} and the estimate of $18.47\pm0.07$~mag by \citet{bhardwaj2016a} using the photometry of Classical Cepheids from Paper I. The error budget includes estimates of uncertainties due to: (1) photometry and extinction corrections (0.05~mag); (2) metallicity corrections, given the uncertainties in [Fe/H] from \citet{benedict2011} and the metallicity correction coefficient uncertainty of 0.07 mag/dex from \citet{muraveva2015}; (3) parallaxes; (4) slope and zeropoint of PLRs. The overall uncertainties are constrained by the inverse weighted variance of the calibrators and the standard deviation of the mean added in quadrature. We note that the LMC distance modulus based on B-W distances is smaller than those based on trigonometric parallaxes, similar to the results of \citet{feast2008}. Therefore, we calculate a mean distance modulus to the LMC based only on the two independent calibrations that rely on parallaxes: $\mu = 18.54\pm0.08$~mag.

\begin{deluxetable*}{llrrrrrrrrr}
\tablewidth{8in}
\tabletypesize{\footnotesize}
\tablecaption{Template-fit mean magnitudes for Type \rom{2} Cepheids in Galactic globular clusters.\label{table:t2c_gcc}}
\tablehead{\colhead{Cluster}&\colhead{Star}&\colhead{P}& \colhead{$\langle J\rangle$}&\colhead{$\langle H\rangle$}&\colhead{$\langle K_s\rangle$}&\multicolumn{3}{c}{$\sigma$}&\colhead{$E_{BV}$}&\colhead{[Fe/H]}}
\startdata
NGC$\,$1904 &  V8& 77.200& 10.346&  9.814&  9.652& 0.064& 0.060& 0.058& 0.01& -1.57\\
NGC$\,$2808 & V10&  1.765& 13.893& 13.513& 13.436& 0.039& 0.033& 0.044& 0.22& -1.15\\
NGC$\,$5139 &  V1& 29.348&  9.373&  9.009&  8.898& 0.052& 0.056& 0.026& 0.12& -1.60\\
NGC$\,$5139 & V29& 14.734& 10.418& 10.015&  9.907& 0.063& 0.066& 0.092& 0.12& -1.60\\
NGC$\,$5139 & V48&  4.474& 11.499& 11.108& 11.023& 0.038& 0.022& 0.075& 0.12& -1.60\\
NGC$\,$5272 &V154& 15.284& 11.348& 11.011& 10.929& 0.044& 0.037& 0.034& 0.01& -1.57\\
NGC$\,$5904 & V42& 25.738& 10.043&  9.697&  9.642& 0.062& 0.072& 0.077& 0.03& -1.27\\
NGC$\,$5904 & V84& 26.870& 10.097&  9.713&  9.625& 0.107& 0.092& 0.093& 0.03& -1.27\\
NGC$\,$5986 & V13& 40.620& 10.912& 10.232& 10.075& 0.024& 0.014& 0.017& 0.28& -1.58\\
NGC$\,$6093 &  V1& 16.304& 11.624& 11.207& 11.090& 0.039& 0.050& 0.038& 0.18& -1.75\\
NGC$\,$6218 &  V1& 15.480& 10.259&  9.780&  9.654& 0.056& 0.027& 0.051& 0.19& -1.48\\
NGC$\,$6254 &  V1& 48.950&  9.083&  8.430&  8.251& 0.049& 0.026& 0.013& 0.28& -1.52\\
NGC$\,$6254 &  V2& 18.723&  9.991&  9.559&  9.416& 0.055& 0.061& 0.059& 0.28& -1.52\\
NGC$\,$6254 &  V3&  7.831& 10.971& 10.537& 10.402& 0.045& 0.020& 0.074& 0.28& -1.52\\
NGC$\,$6256 &  V1& 12.447& 11.766& 11.039& 10.767& 0.081& 0.048& 0.061& 1.03& -0.70\\
NGC$\,$6266 &  V2& 10.609& 11.068& 10.560& 10.409& 0.054& 0.043& 0.065& 0.47& -1.29\\
NGC$\,$6273 &  V1& 16.920& 11.357& 10.873& 10.736& 0.030& 0.028& 0.029& 0.41& -1.68\\
NGC$\,$6273 &  V2& 14.138& 11.466& 11.032& 10.879& 0.038& 0.035& 0.041& 0.41& -1.68\\
NGC$\,$6273 &  V4&  2.433& 13.225& 12.770& 12.684& 0.026& 0.026& 0.043& 0.41& -1.68\\
NGC$\,$6284 &  V1&  4.481& 13.660& 13.223& 13.120& 0.041& 0.038& 0.034& 0.28& -1.32\\
NGC$\,$6284 &  V4&  2.819& 14.111& 13.663& 13.605& 0.037& 0.038& 0.042& 0.28& -1.32\\
NGC$\,$6293 &  V2&  1.182& 14.227& 13.788& 13.648& 0.016& 0.018& 0.060& 0.41& -1.92\\
NGC$\,$6325 &  V1& 12.516& 11.985& 11.275& 11.053& 0.037& 0.038& 0.038& 0.89& -1.17\\
NGC$\,$6325 &  V2& 10.744& 12.131& 11.430& 11.221& 0.014& 0.018& 0.015& 0.89& -1.17\\
NGC$\,$6402 &  V1& 18.743& 11.558& 11.033& 10.834& 0.026& 0.042& 0.048& 0.60& -1.39\\
NGC$\,$6402 &  V2&  2.795& 13.405& 12.954& 12.802& 0.018& 0.025& 0.021& 0.60& -1.39\\
NGC$\,$6402 &  V7& 13.599& 12.035& 11.468& 11.303& 0.033& 0.021& 0.026& 0.60& -1.39\\
NGC$\,$6402 & V76&  1.890& 13.820& 13.351& 13.124& 0.016& 0.009& 0.022& 0.60& -1.39\\
NGC$\,$6441 &V129& 17.832& 12.146& 11.593& 11.471& 0.022& 0.103& 0.068& 0.47& -0.53\\
NGC$\,$6441 &  V6& 22.470& 12.045& 11.599& 11.418& 0.041& 0.088& 0.089& 0.47& -0.53\\
NGC$\,$6453 &  V1& 31.070& 11.470& 10.812& 10.632& 0.037& 0.026& 0.026& 0.66& -1.53\\
NGC$\,$6453 &  V2& 27.210& 11.245& 10.654& 10.482& 0.040& 0.037& 0.019& 0.66& -1.53\\
NGC$\,$6569 & V16& 87.500& 10.502&  9.669&  9.422& 0.105& 0.084& 0.085& 0.55& -0.86\\
NGC$\,$6626 & V17& 48.000&  9.405&  8.809&  8.626& 0.086& 0.075& 0.071& 0.40& -1.45\\
NGC$\,$6626 &  V4& 13.458& 10.757& 10.190& 10.047& 0.055& 0.046& 0.062& 0.40& -1.45\\
NGC$\,$6749 &  V1&  4.481& 13.352& 12.579& 12.323& 0.027& 0.022& 0.019& 1.50& -1.60\\
NGC$\,$6779 &  V1&  1.510& 13.993& 13.631& 13.553& 0.016& 0.040& 0.058& 0.20& -1.94\\
NGC$\,$6779 &  V6& 45.000& 10.711& 10.254& 10.119& 0.026& 0.044& 0.043& 0.20& -1.94\\
NGC$\,$7078 & V86& 16.800& 11.665& 11.261& 11.155& 0.039& 0.035& 0.030& 0.10& -2.26\\
NGC$\,$7089 &  V1& 15.568& 11.939& 11.549& 11.446& 0.018& 0.021& 0.024& 0.06& -1.62\\
NGC$\,$7089 & V11& 33.400& 10.860& 10.479& 10.401& 0.060& 0.037& 0.042& 0.06& -1.62\\
NGC$\,$7089 &  V5& 17.555& 11.803& 11.401& 11.310& 0.032& 0.036& 0.030& 0.06& -1.62\\
NGC$\,$7089 &  V6& 19.360& 11.665& 11.291& 11.204& 0.030& 0.038& 0.042& 0.06& -1.62\\
HP 1        & V16& 16.400& 11.768& 10.975& 10.675& 0.022& 0.019& 0.024& 1.19& -1.50\\
HP 1        & V17& 14.420& 11.872& 11.062& 10.783& 0.044& 0.041& 0.040& 1.19& -1.50\\
Ter 1       &  V5& 18.850& 11.960& 10.921& 10.578& 0.026& 0.027& 0.035& 2.28&\mcr{-1.30}\\
\enddata
\tablecomments{$E_{BV}=E(B-V)$.}
\end{deluxetable*}

\section{Distances to Galactic Globular Clusters}
\label{sec:ggcdist}

\begin{figure}[t]
\begin{center}
\includegraphics[width=0.5\textwidth,keepaspectratio]{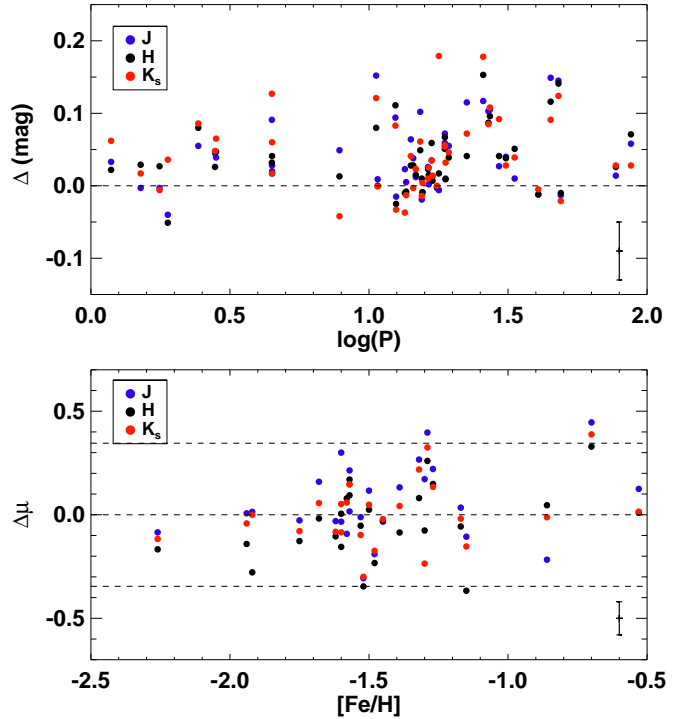}
\caption{Top: difference in mean magnitudes between \citet{matsunaga2006} and template-fit values derived in this work. Bottom: difference in distance modulus obtained by \citet{matsunaga2006} (using the $M_V$-[Fe/H] relation) and this work (using type \rom{2} Cepheid PLRs). Dashed lines indicate $\pm 2\sigma$ of the average difference. Representative median error bar is also displayed in each panel.}
\label{fig:gcc_delm}
\end{center}
\end{figure}

\citet{matsunaga2006} published NIR light curves for T2Cs in 26 Galactic globular clusters and derived the corresponding PLRs. However, the definition of mean magnitude adopted by the authors was a simple mean of maximum and minimum values, which may bias the results since the light curves are neither sinusoidal nor fairly well sampled. Therefore, we use our templates to fit their data and obtain robust mean magnitudes. The resulting light curves are displayed in Fig.~\ref{fig:temp_lc_ggc}, while the mean magnitudes are listed in Table~\ref{table:t2c_gcc}. Fig.~\ref{fig:gcc_delm} displays the difference in mean magnitudes obtained via these two approaches, showing that the results from \citet{matsunaga2006} were significantly biased towards larger values.

In order to obtain distance estimates to these globular clusters, we perform an absolute calibration of the LMC PLRs using the distance modulus derived by \citet{piet13} using late-type eclipsing binaries, $\mu = 18.493\pm0.048$~mag. This estimate is significantly more precise and accurate than the one we obtained in \S~\ref{ssec:lmc_mu} using a few trigonometric parallaxes and was also adopted in Papers \rom{1} and \rom{2} to calibrate the Classical Cepheid PLRs and PWRs.

\begin{deluxetable}{lrrrr}
\tablecaption{Type \rom{2} Cepheid-based distance estimates for Galactic globular clusters\label{table:mu_gcc}}
\tablewidth{0pc}
\tabletypesize{\footnotesize}
\tablehead{\colhead{Cluster} & \colhead{$\mu(J)$} & \colhead{$\mu(H)$} & \colhead{$\mu(K_s)$} & \colhead{$\mu_{M06}$}}
\startdata
NGC$\,$1904 & 15.55$\pm$0.09 & 15.40$\pm$0.09 & 15.42$\pm$0.10 & 15.57\\
NGC$\,$2808 & 15.01$\pm$0.09 & 15.27$\pm$0.09 & 15.05$\pm$0.11 & 14.90\\
NGC$\,$5139 & 13.65$\pm$0.08 & 13.77$\pm$0.08 & 13.70$\pm$0.11 & 13.62\\
NGC$\,$5272 & 14.89$\pm$0.07 & 15.01$\pm$0.07 & 14.95$\pm$0.07 & 15.10\\
NGC$\,$5904 & 14.15$\pm$0.12 & 14.22$\pm$0.11 & 14.24$\pm$0.11 & 14.37\\
NGC$\,$5986 & 15.20$\pm$0.06 & 15.03$\pm$0.06 & 15.05$\pm$0.07 & 15.11\\
NGC$\,$6093 & 15.07$\pm$0.07 & 15.17$\pm$0.07 & 15.12$\pm$0.07 & 15.04\\
NGC$\,$6218 & 13.64$\pm$0.08 & 13.68$\pm$0.06 & 13.62$\pm$0.08 & 13.45\\
NGC$\,$6254 & 13.54$\pm$0.08 & 13.58$\pm$0.07 & 13.53$\pm$0.08 & 13.23\\
NGC$\,$6256 & 14.12$\pm$0.10 & 14.24$\pm$0.07 & 14.18$\pm$0.08 & 14.57\\
NGC$\,$6266 & 13.79$\pm$0.08 & 13.93$\pm$0.07 & 13.86$\pm$0.09 & 14.19\\
NGC$\,$6273 & 14.55$\pm$0.07 & 14.73$\pm$0.06 & 14.65$\pm$0.07 & 14.71\\
NGC$\,$6284 & 15.66$\pm$0.08 & 15.85$\pm$0.08 & 15.71$\pm$0.10 & 15.93\\
NGC$\,$6293 & 14.75$\pm$0.09 & 15.04$\pm$0.09 & 14.76$\pm$0.13 & 14.76\\
NGC$\,$6325 & 14.48$\pm$0.07 & 14.57$\pm$0.07 & 14.53$\pm$0.07 & 14.51\\
NGC$\,$6402 & 14.73$\pm$0.07 & 14.95$\pm$0.08 & 14.82$\pm$0.09 & 14.86\\
NGC$\,$6441 & 15.48$\pm$0.07 & 15.59$\pm$0.12 & 15.59$\pm$0.11 & 15.60\\
NGC$\,$6453 & 14.94$\pm$0.07 & 14.98$\pm$0.07 & 15.03$\pm$0.07 & 14.93\\
NGC$\,$6569 & 15.33$\pm$0.13 & 15.06$\pm$0.11 & 15.12$\pm$0.12 & 15.11\\
NGC$\,$6626 & 13.77$\pm$0.10 & 13.77$\pm$0.10 & 13.76$\pm$0.10 & 13.74\\
NGC$\,$6749 & 14.21$\pm$0.07 & 14.51$\pm$0.07 & 14.46$\pm$0.08 & 14.51\\
NGC$\,$6779 & 15.07$\pm$0.08 & 15.22$\pm$0.09 & 15.12$\pm$0.12 & 15.08\\
NGC$\,$7078 & 15.22$\pm$0.07 & 15.30$\pm$0.06 & 15.25$\pm$0.06 & 15.13\\
NGC$\,$7089 & 15.36$\pm$0.06 & 15.43$\pm$0.07 & 15.41$\pm$0.07 & 15.33\\
HP 1        & 14.24$\pm$0.07 & 14.34$\pm$0.07 & 14.31$\pm$0.07 & 14.36\\
Ter 1       & 13.56$\pm$0.06 & 13.81$\pm$0.06 & 13.97$\pm$0.07 & \mcr{13.73}\\
\enddata
\tablecomments{$\mu_{M06}$: derived by \citet{matsunaga2006} based on\\ magnitudes of horizontal branch stars.}
\end{deluxetable}

We correct the Galactic globular cluster photometry for interstellar extinction using the tabulated $E(B-V)$ values and the \citet{card89} extinction law. We use the absolute calibration of the LMC PLRs to determine distances to each T2C in Table~\ref{table:t2c_gcc} and compute weighted averages for clusters with more than one variable. The error budget includes uncertainties in: (1) mean magnitudes derived from template fits, (2) absolute calibration and (3) eclipsing binary distance to the LMC, added in quadrature. The results are presented in Table~\ref{table:mu_gcc}, along with the estimates by \citet{matsunaga2006}, who used the $M_V-[\rm{Fe}/\rm{H}]$ relation. The bottom panel of Fig.~\ref{fig:gcc_delm} shows the difference between the two approaches; the distances are in agreement within $2\sigma$ in almost all cases.

\vspace{50pt}
\section{Conclusions}
\label{sec:discuss}

We summarize the results of this work as follows:

\begin{itemize}
\item{We present time-series observations of 81 Type \rom{2} Cepheids in the LMC at $JHK_s$ wavelengths, based on the survey of \citet{macri2015}. The $JK_s$ data complements the photometry from the VMC survey \citep{ripepi2015} while the $H$-band time-series observations are presented for the first-time. We develop templates using high-quality and well-sampled light curves of variables in the LMC (observed in $I$-band by OGLE) and the Galactic Bulge (observed in $K_s$-band by VVV).}

\item{We derive robust mean magnitudes based on template fits and obtain Period-Luminosity relations for each class of variable. Our relations are consistent with published work based on variables in the LMC, Galactic globular clusters and the Galactic bulge.}

\item{We compile near-infrared magnitudes for the entire sample of OGLE-\rom{3} Type \rom{2} Cepheids and derive new Period-Wesenheit relations by combining optical and near-infrared data. The slopes of the $W_{V,K_s}$ and $W_{V,J}$ relations are consistent with the findings of \citet{ripepi2015}; in the latter case, when the comparison is restricted to BL Herculis and W Virginis stars.}

\item{We use the {\it Gaia DR1} parallax for VY~Pyx and the {\it HST} parallaxes for $k$~Pav and 5 RR Lyrae variables to obtain an absolute calibration of the zeropoint of the P-L relations. This yields an estimate of the LMC distance modulus of $\mu_{\rm LMC} = 18.54\pm0.08$~mag, in very good agreement with the more accurate and precise estimate by \citet{piet13}. Our estimate is also consistent with recent results based on Classical Cepheids \citep{monson12, bhardwaj2016a}.}

\item{We update the mean magnitudes for Type \rom{2} Cepheids in 26 Galactic globular clusters using our light curve templates and estimate distances to these systems. Our findings are in good agreement with estimates based on horizontal branch stars by \citet{matsunaga2006}.} 
\end{itemize}

\vspace{50pt}
\section*{Acknowledgments}
\label{sec:ackno}
We thank the anonymous referee for his/her valuable comments which improved the content and quality of the manuscript.
We are grateful to Noriyuki Matsunaga, Dante Minniti and Martino Romaniello for reading an earlier version of the manuscript and for useful comments. AB thanks the Council of Scientific and Industrial Research, New Delhi, India, for a Senior Research Fellowship. LMM acknowledges support by the United States National Science Foundation through AST grant number 1211603 and by Texas A\&M University through a faculty start-up fund and the Mitchell-Heep-Munnerlyn Endowed Career Enhancement Professorship in Physics or Astronomy. CCN thanks the funding from Ministry of Science and Technology (Taiwan) under the contract NSC104-2112-M-008-012-MY3. We acknowledge the use of data from the ESO Public Survey program ID 179.B-2002 taken with the VISTA telescope.

\bibliographystyle{apj}
\bibliography{cpapir4}

\appendix

\section{Template-fit light curves}

The three panels of Figure~\ref{fig:temp_lc_lmc} present the result of template fits to the $JHK_s$ light curves of Type \rom{2} Cepheids in the LMC based on data from \citet{macri2015}. Figure~\ref{fig:temp_lc_ggc} shows the same, but for variables in Galactic globular clusters based on data from \citet{matsunaga2006}. Stars are arranged by decreasing period. The $J$ (blue) and $K_s$-band (red) light curves have been offset by +0.25 and -0.5 mag for visualization purposes. $H$-band light curves are shown in violet color. The solid and dashed lines represent the best-fit $I$- and $K_s$-band based templates for each band. The star ``ID'' and ``Period (in days)'' are also provided on the top of each light curve. 

The template-fits are performed using least-square minimization in {\it IDL MPCURVEFIT} routine. Templates can be used to $JHK_s$ light curves with poor phase coverage in order to obtain robust mean-magnitudes. In case of single-epoch observations, accurate amplitude ratios for Type II Cepheids will be required to best-fit observations. For short-period Type II Cepheids, light curves exhibit large scatter with respect to the template-fits, which provide a robust estimate of the uncertainty associated with mean magnitudes.

\begin{figure*}
\begin{center}
\includegraphics[width=1.0\textwidth,keepaspectratio]{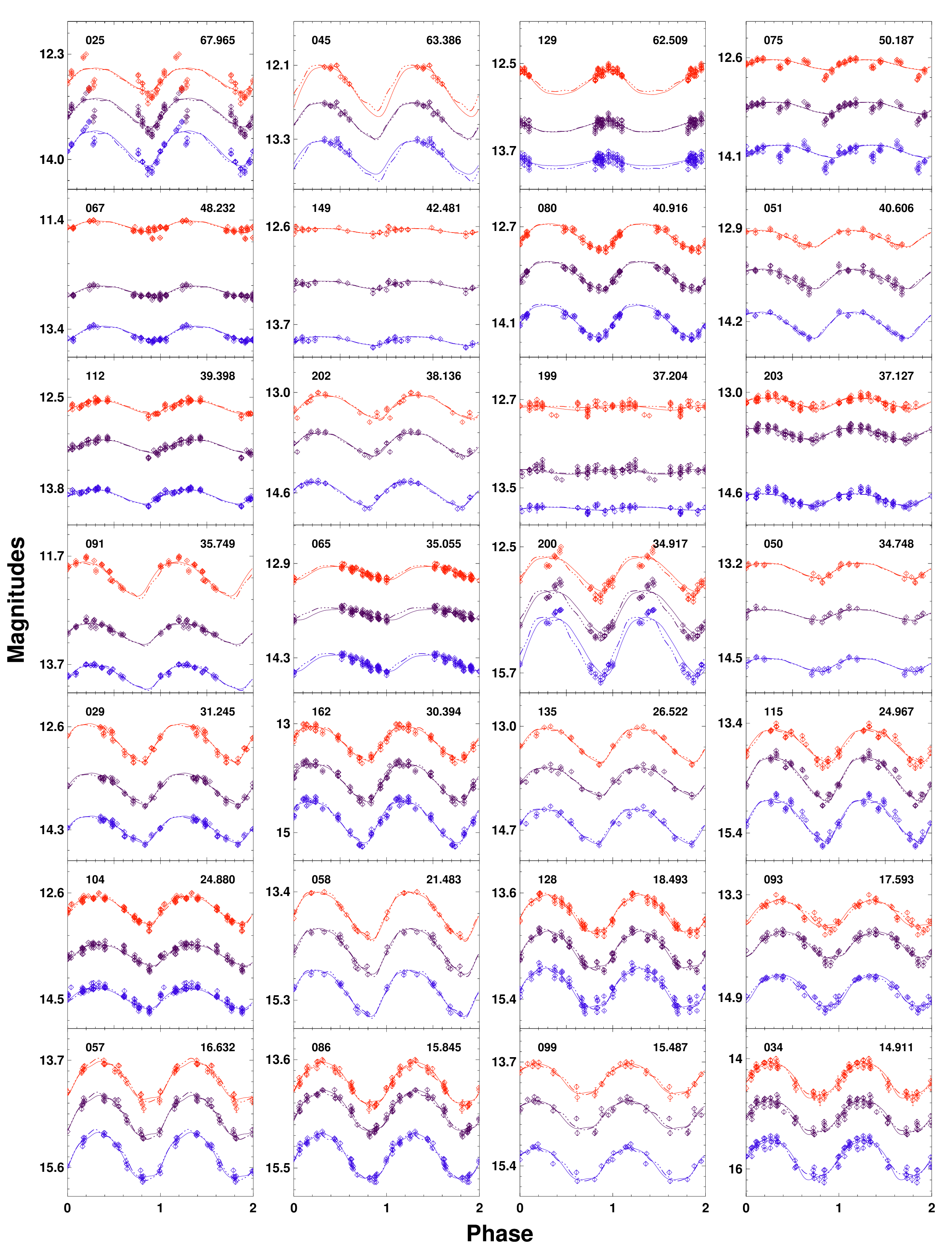}
\end{center}
\caption{Template fits to light curves of Type \rom{2} Cepheids in the LMC, based on data from \citet{macri2015}.} \label{fig:temp_lc_lmc}
\end{figure*}

\addtocounter{figure}{-1}
\begin{figure*}
\begin{center}
\includegraphics[width=1.0\textwidth,keepaspectratio]{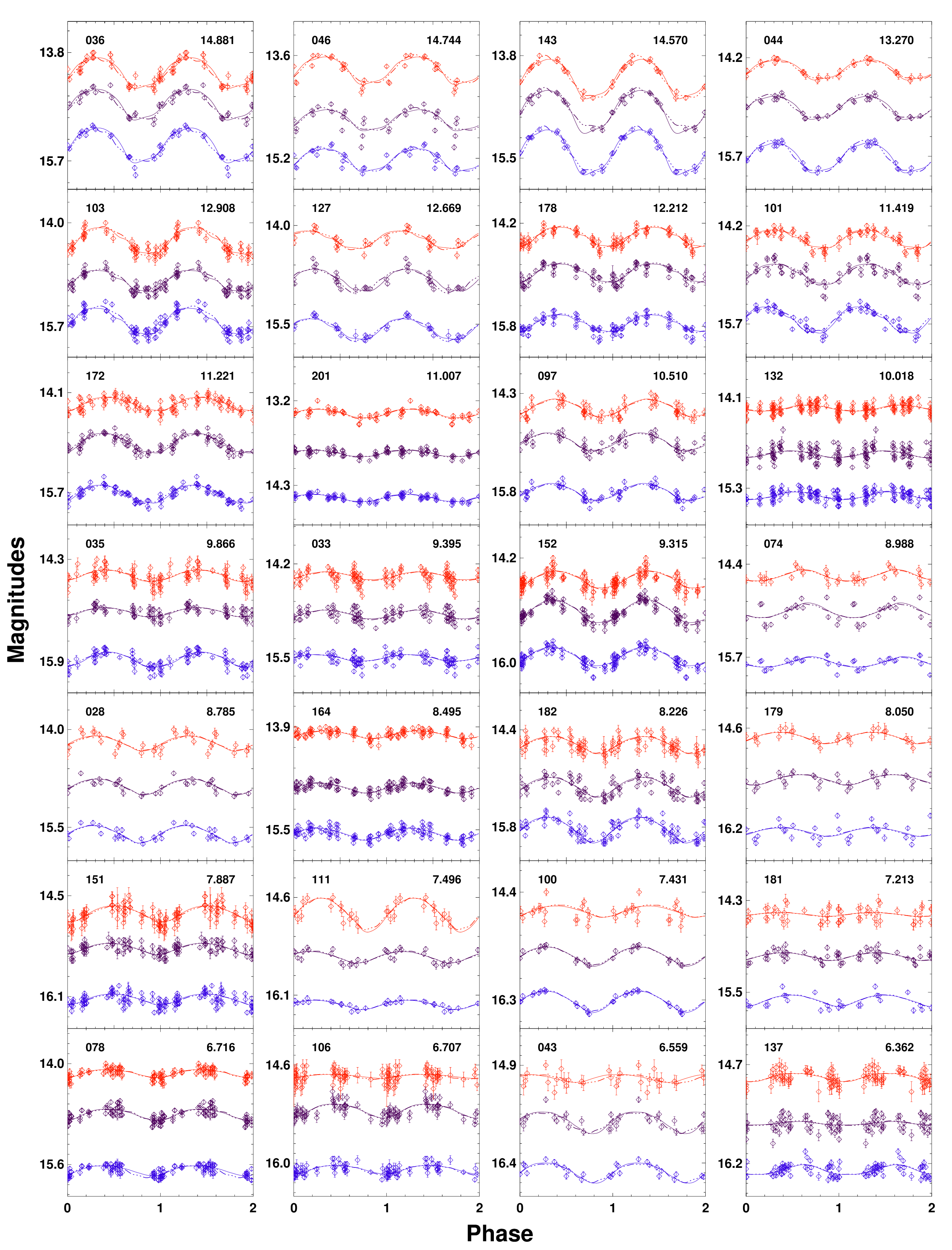}
\end{center}
\caption{(continued).}
\end{figure*}

\addtocounter{figure}{-1}
\begin{figure*}
\begin{center}
\includegraphics[width=1.0\textwidth,keepaspectratio]{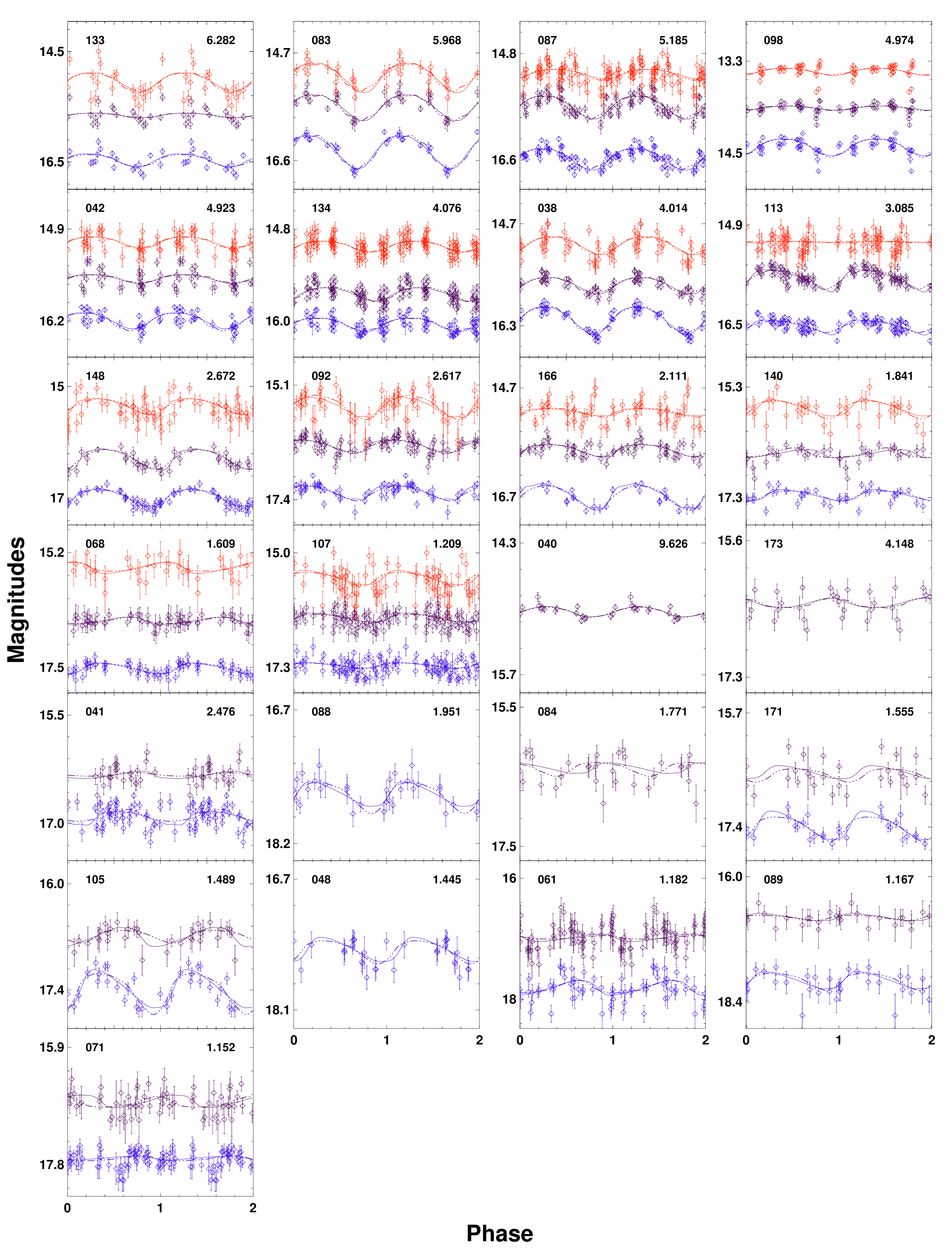}
\end{center}
\caption{(continued).}
\end{figure*}

\begin{figure*}
\begin{center}
\includegraphics[width=1.0\textwidth,keepaspectratio]{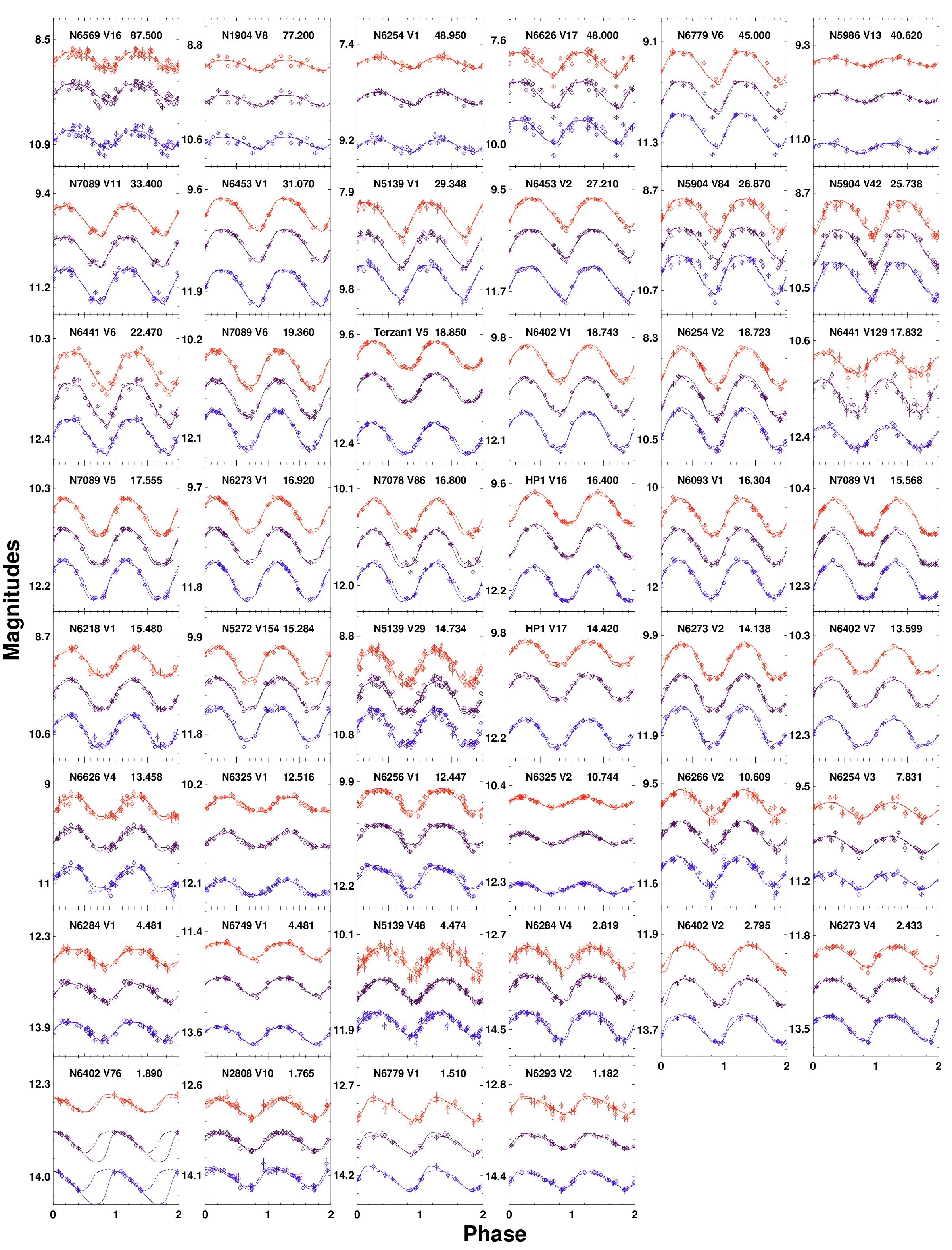}
\end{center}
\caption{Template fits to light curves of Type \rom{2} Cepheids in Galactic globular clusters, based on data from \citet{matsunaga2006}.}\label{fig:temp_lc_ggc}
\end{figure*}

\end{document}